\documentclass[11pt]{article}
\usepackage{amsmath,amssymb}
\usepackage{hyperref}
\usepackage{color}
\usepackage{relsize}
\usepackage{yfonts}
\usepackage{amsbsy}
\usepackage{graphicx}  

\usepackage{makecell}

\usepackage[T1]{fontenc}

\newsavebox{\uuunit}
\sbox{\uuunit}
    {\setlength{\unitlength}{0.825em}
     \begin{picture}(0.6,0.7)
        \thinlines
        \put(0,0){\line(1,0){0.5}}
        \put(0.15,0){\line(0,1){0.7}}
        \put(0.35,0){\line(0,1){0.8}}
       \multiput(0.3,0.8)(-0.04,-0.02){10}{\rule{0.5pt}{0.5pt}}
     \end {picture}}

\newcommand{\ba}{\begin{eqnarray*}}
\newcommand{\ea}{\end{eqnarray*}}
\newcommand{\ban}{\begin{eqnarray}}
\newcommand{\ean}{\end{eqnarray}}

\newcommand{\tr}{{\rm tr\,}}

\newcommand{\IZ}{\mathbb{Z}}
\newcommand{\IC}{\mathbb{C}}

\newcommand{\IH}{\mathbb{H}}

\def\beq{\begin{equation}}
\def\bee{\begin{equation}}
\def\eeq{\end{equation}}
\def\bea{\begin{eqnarray}}
\def\eea{\end{eqnarray}}
\def\bd{\begin{displaymath}}
\def\ed{\end{displaymath}}

\usepackage{tikz}
\newcommand*\circled[1]{\tikz[baseline=(char.base)]{
            \node[shape=circle,draw,inner sep=2pt] (char) {#1};}}

\numberwithin{equation}{section}

\begin{document}


\thispagestyle{empty}
{}


\vskip -3mm
\begin{center}
{\bf\LARGE
\vskip - 1cm
$\boldsymbol{{\cal N}=2}$ Moduli   of AdS$_4$ vacua: A fine-print  study \\ [2mm]}

\vspace{10mm}

{\large
{\bf Constantin Bachas,{\bf $^1$} Massimo Bianchi,{\bf $^2$}
Amihay Hanany,{\bf $^3$} }
 
\vspace{1cm}

{ $^1$ Laboratoire de Physique Th\'eorique de l' \'Ecole Normale
Sup\'erieure,\\
PSL Research University, CNRS, Sorbonne Universit\'es, UPMC
Univ.\,Paris 06,\\
24 rue Lhomond, 75231 Paris Cedex 05, France  \\ 
 [4mm]
$^2$ Dipartimento di Fisica, Universit\`a di Roma  ``Tor Vergata'', and \\
I.N.F.N. Sezione di Roma ``Tor Vergata'',\\
 Via della Ricerca Scientifica, 00133 Roma, Italy
\\
 [4mm]
$^3$ Theoretical Physics Group, Imperial College London\\
 Prince Consort Road, London, SW7 2AZ, UK\\
}}

\vspace{5mm}

\end{center}
\vspace{5mm}

\begin{center}
{\bf ABSTRACT}\\
\end{center}
\smallskip
{We analyze   the  moduli  spaces  near    ${\cal N}=4$  supersymmetric  AdS$_4$
vacua  of string theory by combining and comparing various   approaches:  {\small (a)}
the known exact  solutions  of Type IIB string theory with  localized
 5-brane sourcees;  {\small (b)}   the holographically  dual   3d quiver gauge  theories;
  {\small (c)}  gauged supergravity; and 
  {\small (d)}  the representations  of the 
  superconformal algebra $\mathfrak{osp}(4\vert 4)$. 
Short multiplets containing the marginal ${\cal N}= 2$ deformations  transform  
in the $(2;0)$, $(0;2)$  or   $(1;1)$ representations of the R-symmetry  group
$SU(2)_H\times SU(2)_C $.   The first two  are classified by the chiral rings of the Higgs 
and Coulomb branches, while the latter contain  mixed-branch operators. 
We identify the origin of these moduli in string theory,  
  matching in particular  the  operators of  the chiral rings  with open strings 
  on the  magnetized  5-brane sources. 
Our  results provide  new evidence for  the  underlying  holographic duality. 
 The existence of a large number of bound-state moduli  highlights the 
 limitations of effective supergravity. 
      }

\clearpage
\setcounter{page}{1}

\section{Introduction}

 Along with supersymmetry,  moduli are ubiquitous in string theory.  
 Both must be lifted in the real world,   but  may  manifest themselves at low or intermediate
 energies $\ll {\rm M}_{\rm Planck}$ and  thereby  allow   contact with observations.  Understanding how this happens is
  one of the main themes   of string phenomenology.

  Most earlier  studies employed  either effective supergravities, or  
  gauged supergravities which are consistent truncations  in flux-compactification  backgrounds
  (for a review  of flux compactifications see e.g. \cite{Grana:2005jc,Douglas:2006es,Blumenhagen:2006ci}). 
   One  of our motivations  for  the present work  is to explore the  limitations of these approaches. 
We will consider a class of AdS$_4$  vacua with ${\cal N}=4$ supersymmetry
for which  improved tools are now at our disposal: \smallskip

  {\small (a)}  Ten-dimensional solutions
of Type IIB string theory with fully localized NS5- and D5-brane sources  \cite{Assel:2011xz,Assel:2012cj}; 
 \,{\small (b)} The dual three-dimensional gauge theories based on linear or circular quivers
 \cite{Hanany:1996ie};
 \,{\small (c)}  Corresponding ${\cal N}=4$ gauged supergravities
 \cite{Schon:2006kz,Dibitetto:2011gm,Dibitetto:2011eu,Louis:2014gxa};
 and last but not least  \,{\small (d)} A complete list of unitary representations of the superconformal
 algebra $\mathfrak{osp}(4\vert 4)$ \cite{Dolan:2008vc,Cordova:2016emh}. \vskip 2mm

   \noindent 
   We will 
    provide a  detailed dictionary  between  representations, string excitations,  fields of the quiver theory
    and fields of gauged supergravities, 
identifying in particular the ones  that contain ${\cal N}=2$ supersymmetric moduli.
As  will be  clear,  gauged supergravity retains some,  but not all pertinent information about
the solutions. It  misses  in particular most of the massless moduli. 

     To be sure, these  AdS$_4$ solutions  are far  from  realistic vacua  of string theory.  In addition to a
     plethora of moduli, they also contain  scalar fields with   (mass)$^2 <0$,  though 
     above the Breitenlohner-Freedman stability bound.  One would have to worry about all these modes in any attempt
     to uplift these vacua  to more realistic de Sitter backgrounds. 


    One of the most satisfying  results of our analysis  is the  
    match between string excitations of the solutions
    and chiral fields in the Higgs or Coulomb  branches of the  dual  gauge theories. 
    {This parallels, but is different,  from the analysis in \cite{Assel:2017hck}.}
We will exhibit in particular a selection rule for allowed  representations of the $SU(2)_H\times SU(2)_C$
R-symmetry, whose origin is strikingly different on the two sides.  
This is  a non-trivial test of holographic duality.

\smallskip

     The outline of the paper is as follows:   In section \ref{gravside}  and in   appendix \ref{app1}
     we briefly review the relevant  features of the   Type IIB solutions of \cite{Assel:2011xz,Assel:2012cj}, 
     as well as the  analysis  
       of supersymmetric  vacua in  gauged ${\cal N}=4$ supergravity 
       by Louis and Triendl 
       \cite{Louis:2014gxa}. 
     We then   examine the excitations around these backgrounds, comparing the two approaches. 
     In section \ref{multiplets}  we first review  the  multiplets of  ${\cal N}=4$ superconformal symmetry
     classified in   \cite{Dolan:2008vc,Cordova:2016emh}, 
   and   identify the ones   that contain putative  ${\cal N}=2$ moduli
   (there are no ${\cal N}=4$ moduli,  as  shown  in
   \cite{Cordova:2016xhm}).  We match these multiplets to the excitations of  section \ref{gravside}, summarizing
   our conclusions in a table.  In section \ref{QGT} we 
 introduce the `good' quiver gauge theories of Gaiotto and Witten \cite{Gaiotto:2008ak},   conjectured to flow
 to interacting CFTs that are  dual to the above AdS$_4$ backgrounds.   
We show how  the spectrum of chiral operators on the Higgs branches  of the electric and magnetic quivers
matches nicely with expectations from  string theory,   comment on  mixed-branch operators
and  count  moduli in some examples.  
 
{\it Note added}

In one higher dimension, the similar problem of identifying ${\cal N} =1$ moduli of ${\cal N} =2$
$D=4$ SCFT's with AdS$_5$ holographic duals has been recently addressed in   
\cite{Ashmore:2016oug} from the perspective of generalized exceptional geometry.


\section{ AdS$_4$   vacua of string theory}
\label{gravside}

In this section we review  those features of the ${\cal N}=4$ AdS$_4$
solutions that will be useful later, compare them with the solutions of gauged supergravity, 
and derive some  properties  of the  small-fluctuation 
spectrum. {There is a large literature on AdS$_4$ compactifications with fluxes and branes,
mostly in   Type IIA string theory 
(a non-exhaustive list is    \cite{Kounnas:2007dd}\,-\,\cite{Aharony:2010af}).  
 What is special  about  the solutions of \cite{Assel:2011xz,Assel:2012cj} is that they have fully localized 
 (as opposed to smeared)  brane sources. A   general method of searching  for such  solutions
with the help of pure-spinor equations   has been proposed recently in
 \cite{Passias:2017yke}. 
   
 }


\subsection{Review  of the  $\boldsymbol{{\cal N}=4}$ Type IIB solutions}

We begin with the ${\cal N}=4$ AdS$_4$ compactifications of
Type IIB
string theory found in \cite{Assel:2011xz,Assel:2012cj}.
The geometry of these solutions
has the warped form (AdS$_4\times$S$^{2}\times$S$
^{2\,\prime}$)$\times_w\Sigma$,
where the base $\Sigma$ is an open Riemann surface, which is
either the disk or the annulus.
Superconformal symmetry $\mathfrak{osp}(4\vert 4)$ is realized as 
isometries of the AdS$_4\times$S$^{2}\times$S$ ^{2\,\prime}$
fiber. R-symmetry, in particular, is realized as isometries of the
two 2-spheres.
         
In addition to the metric, the solutions feature non-trivial
dilaton, 5-form and complex 3-form backgrounds.
As shown by D'Hoker, Estes and Gutperle
\cite{D'Hoker:2007xy,D'Hoker:2007xz} all these  backgrounds  can  be 
expressed in terms of two 
harmonic functions $h_i: \Sigma \to \mathbb{C}$ $(i=1,2)$, which
are positive in the interior
of $\Sigma$ and vanish  at alternating parts of the boundary. This  latter
property ensures that
generic points of $\partial\Sigma$ are  regular
interior points of the 10-dimensional geometry. 
The explicit expressions of all  background fields are given in appendix \ref{app1}\,.  


A key feature of the solutions is that they have localized
D5-brane and NS5-brane sources (at finite distance)  on the  internal space 
${\cal M}_6 := ($S$^{2}\times$S$ ^{2\,\prime})\times_w\Sigma$\,.
The D5-branes wrap the S$^2$ fiber and are localized in the
transverse space S$^{2\,\prime}\times_w\Sigma$,
whereas the NS5-branes wrap the S$^{2\,\prime}$ fiber and are
localized in S$^2\times_w\Sigma$.  These cycles are homologically trivial, so 
there is no need for  tadpole cancellation
by orientifolds or anti-branes. 

We focus on the solutions with  $\Sigma$  the infinite strip $0\leq {\rm Im}z \leq \pi/2$
(the case of the annulus is a simple extension). The  
 singularities  labelled by $a\in \{ 1, \cdots , p\}$\ have  D5-brane  charge   $N_a$ and    are 
   located at   Re$z  = \delta_a$
on   the upper strip boundary, while the   
 singularities
   labelled   by $\hat a  \in \{ 1,  \cdots , \hat p\}$\, have NS5-brane  charge $\hat N_{\hat a}$ and position
   Re$z= \hat\delta_{\hat a}$ on the lower  boundary.  Both the 5-brane charges and their positions are 
   continuous parameters of the supergravity solutions 
    but in  string theory they are all quantized. 
    This  is obvious for the 5-brane charges, but  more subtle for their positions $\{\delta_a , \hat \delta_{\hat a}\}$.  
    It turns out that these latter  can be  related to the D3-brane charges of the 5-brane stacks  \cite{Assel:2011xz}
\bea
\ell_a = -  \sum_{{\hat a}=1}^{\hat p}\hat N_{\hat a}\, {2\over \pi} \arctan
(e^{- \delta_a+ \hat\delta_{\hat a} })\ , 
\qquad 
\hat \ell_{\hat a} =   \sum_{a=1}^{ p} N_a \, {2\over \pi} \arctan
(e^{- \delta_a+ \hat\delta_{\hat a} })\ , 
\eea
where $\ell_a$ is the D3 charge of  a D5-brane in the $a^{\rm th}$  stack  and
$\hat \ell_{\hat a}$  the D3 charge of a NS5-brane in the ${\hat a}^{\rm th}$   stack.
The above   equations can be used to solve for   all  source positions  in terms of the  charges 
$\{ N_a, \ell_a, \hat N_{\hat a}, \hat \ell_{\hat a}\}$
which are quantized.

    Thanks to this  dimensional transmutation (charges transmuting to geometric positions)
the solutions of \cite{Assel:2011xz,Assel:2012cj} have only discrete but no continuous moduli. 
As we will discuss  in section \ref{multiplets}\,,  
the absence of continuous ${\cal N}=4$ superconformal moduli follows  in all generality
from  the study of unitary $\mathfrak{osp}(4\vert 4)$ representations \cite{Cordova:2016xhm}.  
Charge quantization removes therefore a potential contradiction with this general result. 


   The  data $\{ N_a, \ell_a\}$ and $\{\hat N_{\hat a}, \hat \ell_{\hat a}\}$
   can be repackaged conveniently   in two Young diagrams   $\rho$ and $\hat\rho$.
 The    diagram
  $\rho$ has $N_a$ rows of $\vert \ell_a\vert $ boxes  (with the $\vert \ell_a\vert $ arranged in descending order). 
   Likewise    $\hat \rho$ has $\hat N_{\hat a}$ rows of $\hat \ell_{\hat a}$ boxes. 
   In  the  Type IIB  solutions  
  $\rho$ and $\hat\rho$   are not independent:  
  they   have the same total number of boxes by  conservation of D3-brane charge 
(one can indeed check that $ \sum_{\hat a} \hat N_{\hat a} \hat \ell_{\hat a} =  -\sum_a N_a\ell_a :=  N$),   
 and they  furthermore automatically  satisfy 
  the  partial ordering condition 
   \bea\label{order}
    \rho^T  > \hat\rho\ .
   \eea
   Here $\rho^T$ is the transposed Young diagram (with columns and rows interchanged)
   and the above condition means that the first $k$ rows of the left diagram  contain more boxes than
   the first $k$ rows of  the right diagram,  for all 
   $k=1, \cdots , \hat p-1$.\,\footnote{This condition has appeared earlier   in  related contexts
   \cite{Kronheimer:1990ay}\cite{Bachas:2000dx}.
   It is also natural in the Brieskorn-Slodowy theory of transverse slices  \cite{Bri,Slo} 
   reviewed in  section 4.
        It would be interesting to see if this condition  can be
   obtained from some appropriate version of K theory along the
    lines of \cite{Kapustin:1999di}\cite{Bouwknegt:2000qt}.  } 
   In the dual linear-quiver gauge theory,  where $\ell_a$, $\hat\ell_{\hat a}$ become the linking numbers 
   of the 5-branes, 
    this  condition  ensures that the gauge symmetry can have a non trivial Higgs branch, see  section \ref{QGT}\,.
   Such  theories were 
  called `good'  by Gaiotto and Witten \cite{Gaiotto:2008ak} and conjectured  to flow to 
   strongly-interacting    fixed points in the infrared. 
  That the same conditions also arise   in string theory is a  nice
  check of holographic duality \cite{Assel:2011xz}. 
 
 The gauge symmetries on the worldvolumes of the 5-branes
correspond
to global  flavor symmetries of the  dual field
theory.
 The D5-brane symmetry  $U(N_1) \times \cdots \times U(N_p)$  
  is  manifest when the CFT is realized as
     an  `electric  quiver' gauge theory, 
while the NS5-brane symmetry  $U(\hat N_1) \times \cdots \times U(\hat N_{\hat p})$
is  manifest in the {Lagrangian of the}  mirror
`magnetic quiver'.
  At the origin of the Higgs branch, the electric-  and magnetic-quiver 
 gauge
       theories are expected  to flow to the same CFT  where 
    these  flavor  symmetries  coexist.

 \begin{figure}[t!]
 \vskip -11mm
\centering 
\includegraphics[width=.72\textwidth,scale=0.60,clip=true]{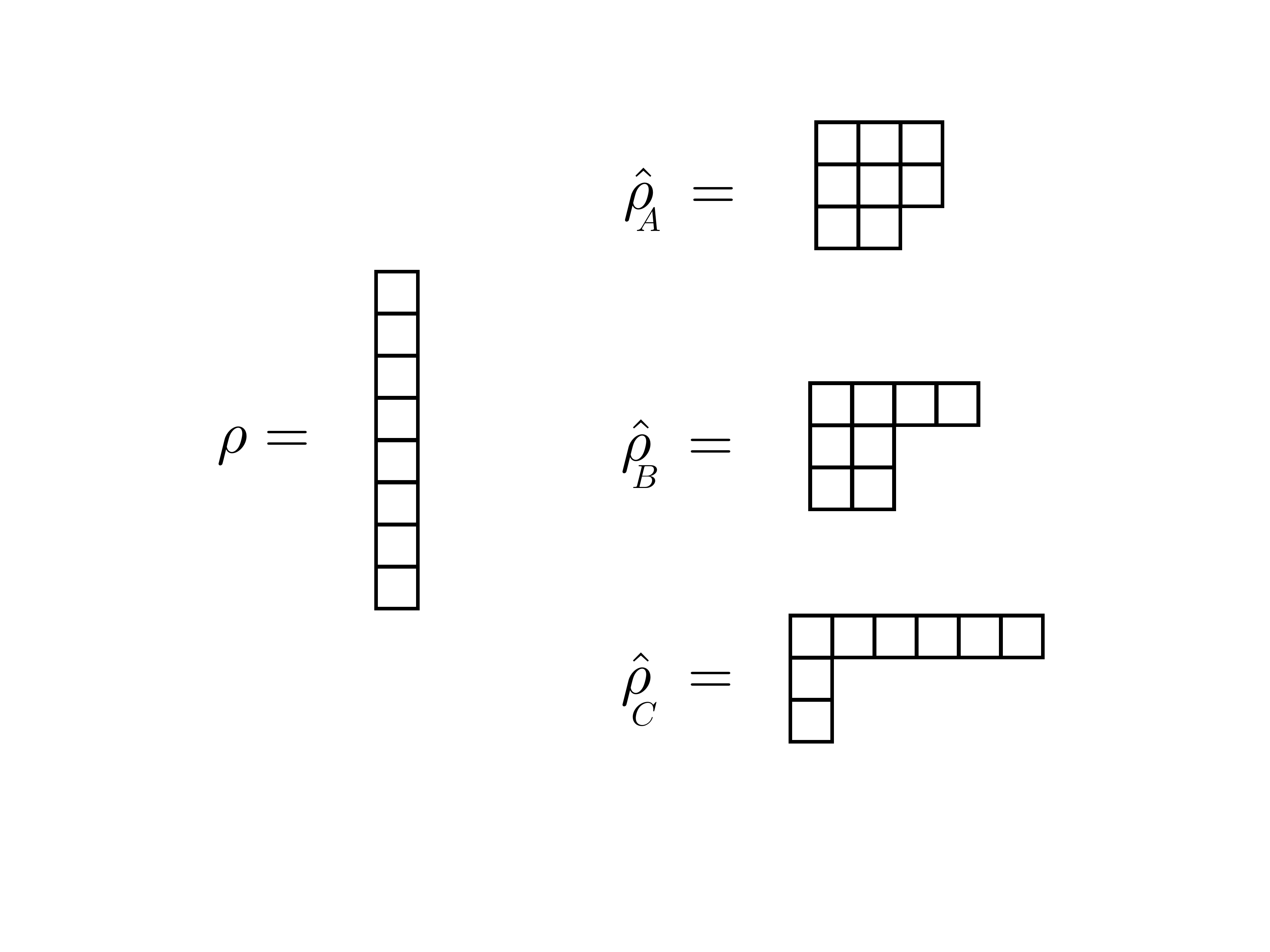}
  \vskip -14mm
   \caption{\small Three   theories $(A,B,C)$  with  the same $\rho$ and different $\hat\rho$.  These theories admit the same continuous 
   global symmetry $SU(8)\times \bigl(U(2)\times U(1)\bigr)/U(1)$  but different fine print. The gravitational dual solutions all consist of one stack of $N=8$  D5-branes and two stacks of  $\hat N_1= 2$ and $\hat N_2=1$ NS5-branes, but they  differ
   in the distribution of the 8 units of  D3-brane charge. }
\label{fig:1} 
 \end{figure}

       It is important here to note    that  
        global symmetries  do  not determine the SCFT uniquely.
        There is  extra information contained in the D3-brane charges or linking numbers  
 $\{\ell_a, \hat \ell_{\hat a}\}$  of the 5-branes. 
  We will refer to this extra information as the `fine print'. 
       Figure \ref{fig:1}  shows  an example of three
       theories with the same 5-brane charges $\{N_a, \hat N_{\hat a}\}$, and hence the
       same global  flavor  symmetry,\,\footnote{In principle some theories could be distinguished by 
       extra discrete symmetries. 
       This does  not seem to be the case for the examples  of the figure, {even though
       we cannot rule out  the possibility that some higher-form discrete symmetry  emerges  in the infrared.
       Note in passing that a diagonal electric and a diagonal magnetic $U(1)$   act  trivially and
       can be modded out of the global-symmetry groups.} 
       but with different fine print. 
        These theories have  different  operator spectra and AdS$_4$  duals as will become clear
      later on. 
  }


 \subsection{Gauged    supergravity}
 \label{gaugesugra}

        A different approach to supersymmetric AdS$_4$ vacua  is the approach   of gauged supergravity \cite{deWit:2005ub, Schon:2006kz, Dibitetto:2011gm, Dibitetto:2011eu,Louis:2014gxa}.\,\footnote{There is a vast  literature on gauged supergravity 
        and stabilization
        of moduli by fluxes,  the above references are just  closer to the contents of this paper. 
         A recent review is  \cite{Trigiante:2016mnt}.
        }
        For   the case at hand one starts with  
         ${\cal N}=4$ supergravity coupled to  $m$ vector multiplets in 4d Minkowski spacetime, and then deforms this  theory
         by gauging a subgroup of the global  symmetry   group SL$(2)\times$SO$(6,m)$.  Gauging is elegantly
         achieved by the  introduction of  an embedding tensor that  obeys   a set of quadratic constraints \cite{deWit:2005ub}.  
        The result is not an effective low-energy theory, but its solutions are guaranteed to be solutions of any 
        higher-dimensional theory of which the gauged  supergravity is a  consistent truncation. 

                                The existence of  maximally-supersymmetric AdS$_4$ vacua  has been  studied
                                systematically 
                                 within this formalism  by Louis and Triendl \cite{Louis:2014gxa}.  
                                  Before gauging, the scalar fields of the  ${\cal N}=4$ supergravity  take values in the coset  space
              ${\cal M} =  [{{\rm SL}(2)/ {\rm SO}(2)}]\times [{{\rm SO}(6,m)/ {\rm SO}(6)}\times SO(m)]$. 
Gauging generates a potential that  lifts some of these flat directions.  Louis  and Triendl showed that whenever maximally supersymmetric AdS$_4$ vacua exist the following is true:
\begin{itemize}
                                \item  The gauge group 
                                    is of the  form  G$_+\times $G$_-\times $G$_0^v$, 
                                where G$_\pm \subset $ SO$(3, m_\pm)$  
                                and G$_0^v\subset$SO$(q)$ with $m_++m_-+q=m$.  
                                Furthermore,  the gauging of G$_+$ employs the electric components of  three
                                graviphotons while the gauging of G$_-$ employs the magnetic components of  the other three graviphotons.  
                                \item  The  vacua break spontaneously  G$_\pm$ to their maximal compact subgroups
                                SO(3)$_\pm \times H_\pm$ where  $H_\pm\subset$ SO$(m_\pm)$. The only 
                                continuous moduli 
                                   are the Goldstone bosons of  these broken   gauge symmetries that
                                are eaten by the corresponding massive vector bosons and are hence unphysical.
                                 As a result, there are no
                                ${\cal N}=4$ preserving continuous moduli. 
             \end{itemize}
                             
        Let us  compare these conclusions   with the findings of the previous subsection.  
         Clearly SO$(3)_+\times$SO$(3)_-$  is   the ${\cal N}=4$ R-symmetry and it is 
         natural to identify  $H_+$ and $H_-$   with the flavor symmetries realized,  respectively,  on
          D5-branes and  NS5-branes. The extra factor G$_0^v$ can be a priori attributed to either type of 5-branes, 
          or may come from  extra bulk vector bosons.  The absence of continuous moduli in  this description   is  remarkable.
          For instance   SL$(2, R)$ transformations of Type IIB  supergravity which activate the 
           RR axion field  take   us outside the class of solutions given in  appendix \ref{app1}\,. 
          Such  transformations can generate new (`orbifold equivalent') solutions \cite{Assel:2012cj},
            but they are
          discretized by the fact that all  5-brane charges   must remain   integer.
          The fact   that gauged supergravity has no continuous SL$(2, R)$ moduli shows that 
             the embedding-tensor formalism   knows about integrality of charges.

        Of course 
       gauged supergravity cannot know everything
        about the solutions,   since it is a truncation of the full string theory. 
   It is   unclear, in particular,     whether it can store the `fine print' data
          of the previous subsection, i.e. distinguish  vacua with the same  unbroken gauge symmetry.
           In gauged supergravity this fine print can only come from  inequivalent embeddings
          $H_+\times H_-\times$G$_0^v\subset\,$SO$(6, m)$,   or from  multiple solutions of the quadratic constraints. 
          In many previous  studies  
          the number of vector multiplets was  small  (usually $m=6$) so  the possibilities were  restricted. 
          The  question  deserves further scrutiny.

        What is certain is that  gauged supergravity 
        misses many  scalar  excitations including some   with (mass)$^2\leq 0$ which correspond to
        marginal and relevant deformations of the dual CFT.  
        Apart from   the multiplets of the graviton and the  vector bosons of unbroken gauge symmetries,
        the only other  surviving fields   are  the multiplets of massive gauge bosons 
        corresponding to   the broken  
        non-compact gauge symmetries. 
        These are  in  representations  $(1;0)$ or $(0; 1)$ of the R-symmetry group, i.e. they carry a vector
        index under either SO$(3)_+$ or SO$(3)_-$.\footnote{For illustration,  consider  gauging  
         G$_+$=SO$(3,2)\subset$SO$(3,m_+)$ where $m_+=7$. 
        Note that 3+7 = 10 is precisely the dimension 
        of the gauge group.  The  maximal compact subgroup of this latter is SO$(3)\times$U(1). The six 
        non-compact generators
         transform  in   the $(3, \pm)$ representations of this unbroken gauge group.}
        We will soon see    that they belong
        (in the notation of \cite{Cordova:2016emh},  modified only by halving the spins)
          to the superconformal multiplets 
        $B_1[0]_2^{(2;0)}$ or  $B_1[0]_2^{(0;2)}$  
         which provide  some, but not all, of the ${\cal N}=2$  superconformal moduli.
          Thus   gauged supergravity   is not
           a reliable tool for addressing  the problem of moduli stabilization.

                
   \subsection{Properties of vacuum  excitations}
   \label{excitations}                
      
      The complete spectrum of small excitations can be  in principle derived by expanding around  the
      backgrounds of   \cite{Assel:2011xz, Assel:2012cj}\,.  
      In  practice this is a  formidable  task.  We will  limit ourselves  to  some  generic features that  are
      easy  to  extract. 
      
         We will use certain  facts   about  representations of the superconformal algebra   
      $\mathfrak{osp}(4\vert4)$, postponing a   more systematic discussion to  the following section. 
      Unitary  representations are   decsribed by the spacetime spin of the  highest-weight state, 
      and by its   $SO(3)_+ \times SO(3)_-$ R-symmetry spins
      which are denoted   $(R;R^\prime)$.  When not otherwise qualified, `spin'  means  spacetime spin. 
      Apart from the long  representation $L[0]_1^{(0;0)}$ which corresponds to a massive supergraviton, 
      there are  three series of representations with maximum spin $\leq 2$: \cite{Cordova:2016emh}
      \begin{itemize} 
      \setlength{\itemindent}{3em}
       \item the $B_1[0]_R^{(R;0)}$ and  $B_1[0]_{R^\prime}^{(0;R^\prime)}$ series\ with\ max. spin \ $\leq 1$,  
      \item  the  $B_1[0]_{R+R^\prime}^{(R;R^\prime)}$ series  ($RR^\prime \not=0$) \ with\ max. spin \ $\leq 3/2$,
      \item  the  $A_2[0]_{R+R^\prime+1}^{(R;R^\prime)}$ series\ with\ max. spin \ $\leq 2$. 
      \end{itemize}
      Barring  excited-string modes,   
       single-particle states   of the  theory  are either
      the   10d graviton multiplet,  or  the lowest-lying modes  of  
      open strings   living   on the    5-branes.  Both have spins not exceeding $2$,  and are  therefore
     organized in the above  representations.

                   Consider first the open strings   which 
     contain spins $\leq 1$ and belong  to the  $B_1[0]_R^{(R;0)}$ series. 
            We  concentrate on the   D5-branes, the discussion of NS5-branes is mirror symmetric. 
             Strings  on the $a^{\rm th}$ 
    D5-brane stack  
    transform in the adjoint representation of $U(N_a)$, while those 
     stretching from the $a^{\rm th}$ to 
    the $b^{\rm th}$ stack transform in the bifundamental $(N_a, \bar N_b)$  representation  of $U(N_a)\times U(N_b)$. 
     {There are no open strings stretching between  D5-branes and  NS5-branes, so  CFT 
    operators charged under both  
    electric and  magnetic flavor groups  can only  correspond to multi-string  states. }
             
              We now focus  on the  spin-1 component  of the multiplet  which transforms  in the $(R-1  ;0)$ 
    representation  of  $SO(3)_+$, 
    see eq.\,\eqref{b1mult} of  the following section.  Clearly  $R-1$ is the angular momentum of the string on
    S$^2$, the 2-sphere wrapped by all  the D5-branes. 
    We  can constrain the 
      range  of $R$ for D5-brane strings   by the following argument: 
    A         D5-brane  of the $a^{\rm th}$ stack  carries   $\ell_a$ units of D3-brane charge,
    which must be induced by  $\ell_a$ units
    of  internal  monopole   flux   in the 2-sphere direction \cite{Douglas:1995bn}. 
    Therefore open strings from  $a$ to  $b$ 
   feel   
   a  monopole field of  strength $\ell_a -   \ell_b$ in appropriate units. 
    Their  spin-1 components are classified 
    by the scalar 
   monopole harmonics  on S$^2$,    which have    
   $R-1\geq {1\over 2} \vert \ell_a - \ell_b \vert$,  see e.g.   ref.\,\cite{Weinberg:1993sg}. 
   This leads then to the following selection rule:
   \bea
    R =  {1\over 2} \vert \ell_a - \ell_b \vert + n
    \qquad (n=1, 2, \cdots)  \qquad {\rm for}\ \,  (ab) \ \, {\rm open \ strings}\ . 
   \eea
   We will see that this  prediction agrees beautifully with  the analysis  of  the  CFT side. 
   It is important also to note  that  the S$^2$ angular momentum can be half-integer, a well-known fact  of physics 
   in the background of magnetic poles \cite{Jackiw:1975fn, Hasenfratz:1976gr}.

    \smallskip
    
        We come next to the closed-string modes.  Barring again string excitations, 
     these include the  ${\cal N}=4$ supergraviton in four dimensions, its Kaluza Klein excitations, 
    and massive spin-3/2 or  spin-1 
    multiplets  corresponding to the  supersymmetries and R-symmetry generators     broken
    by the compactification. All closed strings are of course  flavor singlets. 
    In addition, since 10d closed-string states  are
either tensors or spinors on
all three (pseudo)spheres of the
AdS$_4\times$S$^2\times$S$^{2\,\prime}$ fiber simultaneously, the quantum
numbers  $R$ and $R^\prime$ of the highest-weight  state (which is always scalar)
 must be 
integer.\,\footnote{There can   be an  exception to this  rule   if  a 2-sphere is  transpersed
by magnetic  flux of a gauge  field under  which the closed string is  electrically  charged. This situation does not
arise  in the solutions of interest. 
} These simple facts will also emerge automatically on the CFT side. 
    
     Gauged supergravity retains three types of fields:  {\small (i)} the massless  4d  graviton coupling  to the stress tensor
    multiplet,    $A_2[0]_1^{(0;0)}$  in the notation of
    ref.\,\cite{Cordova:2016emh};  {\small (ii)} the massless vector bosons of compact symmetries  coupling to
    conserved-current multiplets $B_1[0]_1^{(1;0)}$ or $B_1[0]_1^{(0;1)}$, and 
    {\small (iii)}  the massive vector bosons of broken non-compact symmetries, which couple to  the 
      $B_1[0]_2^{(2;0)}$ or $B_1[0]_2^{(0;2)}$ multiplets (whose spin-1 components are vectors of $SO(3)_+$ or
      $SO(3)_-$).  We will see that these last multiplets contain marginal ${\cal N}=2$ supersymmetric 
      operators, but they are not the only ones.   Marginal ${\cal N}=2$  operators are also contained in the 
      spin-3/2 multiplet    $B_1[0]_2^{(1;1)}$ which  is truncated out in  gauged ${\cal N}=4$  supergravity. 
      We expect  such multiplets  to  exist in half-supersymmetric vacua of ${\cal N}=8$  supergravity,
      where four  of the  gravitini become massive, but have not checked this explicitly.

   An example of a deformation  that excites the   truncated modes   is the TsT deformation of 
      Lunin and Maldacena \cite{Lunin:2005jy}. For the backgrounds of interest, the two 
      commuting Killing isometries  are the azimuthal rotations of S$^2$ and S$^{2\,\prime}$. 
      If  $\tau$ is the complexified K\"ahler modulus of the  2-torus generated by these Killing isometries,
      the deformed solutions have 
\bea\label{TsT}
\tau \rightarrow \tau^\prime = {\tau\over 1+\gamma\tau}\ .  
\eea
where  $\gamma$ is the real deformation parameter.
Besides the metric and $B$-field components that are encoded in   $\tau$, several other supergravity fields 
 are  also deformed -- 
  explicit general formulae  have been worked out by Imeroni \cite{Imeroni:2008cr} 
  but are not  needed  here. One crucial  remark   is that in the original backgrounds $B$ has
  no 2-torus component, so $\tau = B + i \sqrt{G}$ is purely imaginary.  
 This guarantees that the deformed solutions do not develop singularities at the loci where
 the torus degenerates \cite{Lunin:2005jy}.
  
               The  TsT deformation  breaks the $SO(4)$ R-symmetry to $U(1)\times U(1)$ which is 
             compatible with at most ${\cal N}= 2$ supersymmetry.  Generally-speaking  we  expect  all four supersymmetries
             to be broken by this deformation\,\footnote{In the dual field theory the  TsT  deformation
             inserts    charge-dependent phases in front of   field products, $fg \to  \exp(i\gamma (Q_1(f)Q_2(g)- Q_1(g)Q_2(f))$
             where $Q_1, Q_2$ are the two $U(1)$ charges being used.  When one of these charges    is the R-charge,  
              different components in  the  product  of two superfields   acquire different phases. 
              For general charges  supersymmetry is thus broken.}\,.  
              The point we want to make  however  here is different: this is a deformation  of the classical background
              that is truncated away in the gauged ${\cal N}= 4$ supergravity  description. Indeed, this deformation is 
                generated by a  neutral scalar operator 
             of scaling dimension $\Delta = 3$ in the dual CFT, and there is  no such operator in the ${\cal N}= 4$ stress-tensor
             multiplet  $A_2[0]_1^{(0;0)}$.


\section{Superconformal Multiplets}
\label{multiplets}

The ${\cal N}=4$ superconformal algebra is $\mathfrak{osp}(4\vert 4)$.  Its maximal 
compact subalgebra includes spatial  rotations and the R-symmetry algebra  
${\mathfrak{su}}(2)\oplus
 {\mathfrak{su}}(2)^\prime \simeq
{\mathfrak{so}} (4)_R $.
Unitary representations of this algebra have been classified,  see 
  Dolan \cite{Dolan:2008vc}, 
and Cordova, Dumitrescu and Intriligator who classified representations of superconformal algebras in all dimensions
 \cite{Cordova:2016emh}. 
 
 We adopt  the
conventions of this  latter   work
and   denote the highest-weight states in a given  
representation by
$[j]_\Delta^{(R; R^\prime)}$, where $\Delta$ is the scaling
dimension and
$(j, R, R^\prime)$ are   the spins   
with respect to space rotations and the two  R-symmetry factors.
 Contrary to \cite{Cordova:2016emh}, our spins
will be however canonically normalized. 
For instance $[1]_\Delta^{(0; 1)}$ is a vector field in the
(singlet, vector) representation of
R symmetry. When this is a superconformal primary, the ${\cal N}
=4$ representation built on it
  is denoted as in reference \cite{Cordova:2016emh},  e.g.   
  $A_1[1]_3^{(0; 1)}$  for a short representation in the 
  $A_1$ series, or\  $L[1]_{\Delta\geq 3}^{(0; 1)}$ for a long representation. 

We start with a short review of superconformal multiplets,  their spin content and matching string-theory modes, 
 then decompose them  under ${\cal N}=2$ supersymmetry 
and identify the ones that contain    ${\cal N}=2$ moduli.


\subsection{ $\boldsymbol{{\cal N}=4}$   multiplets}
\label{3.1}
 
 The 8 independent Poincar\'e 
supercharges can be labelled by the projections of the three
spins, $Q_\pm^{(\pm; \pm)}$.
The primary conformal operators are obtained by acting with
these supercharges
on the primary superconformal operators. There exist 128 bosonic
and 128 fermionic raising operators
transforming as follows under ${\mathfrak{su}}(2)_{\rm
space}\oplus [{\mathfrak{su}}(2)\oplus
{\mathfrak{su}}(2)^\prime]_{\rm R-symmetry}$ :
\bea
1 \, (1): && \qquad \qquad \qquad \qquad \qquad (0,0,0)
\nonumber \\
Q \, (8) : && \qquad  \qquad \qquad  \qquad \quad  \hskip 1.9mm
\textstyle{ (
\scalebox{0.9 }{ 
   ${1\over  2}  ,  {1\over  2}  ,  {1\over  2}$ 
   } } ) 
   \nonumber \\
Q^2 \, (28) :&& \qquad \qquad \qquad \Bigl[ (1, 1, 0) \oplus
{\rm perms} \Bigr] \oplus (0,0,0) \nonumber \\
Q^3  \, (56) :&&  \qquad  \qquad \quad  \Bigl[
\textstyle{ (
\scalebox{0.9 }{ 
   ${3\over  2}  ,  {1\over  2}  ,  {1\over  2}$ 
   } } ) 
 \oplus  {\rm perms}\Bigr]  \oplus 
\textstyle{ (
\scalebox{0.9 }{ 
   ${1\over  2}  ,  {1\over  2}  ,  {1\over  2}$ 
   } } )  
  \nonumber \\
Q^4 \, (70) :&& \quad \Bigl[ (2, 0, 0) \oplus {\rm perms}\Bigr]
\oplus (1,1,1) \oplus \Bigl[ (1, 1, 0) \oplus {\rm perms}\Bigr] \oplus
(0,0,0) \nonumber \\
Q^5 \, (56) :&&  \qquad   \qquad \quad   \Bigl[ 
\textstyle{ (
\scalebox{0.9 }{ 
   ${3\over  2}  ,  {1\over  2}  ,  {1\over  2}$ 
   } } )  
\oplus   {\rm perms}\Bigr]  \oplus
 \textstyle{ (
\scalebox{0.9 }{ 
   ${1\over  2}  ,  {1\over  2}  ,  {1\over  2}$ 
   } } )  
    \nonumber \\
Q^6 \, (28) :&& \qquad \qquad \qquad \Bigl[ (1,1, 0) \oplus {\rm
perms} \Bigr] \oplus (0,0,0) \nonumber \\
Q^7 \, (8) : && \qquad \qquad \qquad \qquad \quad \hskip 1.9mm
\textstyle{ (
\scalebox{0.9 }{ 
   ${1\over  2}  ,  {1\over  2}  ,  {1\over  2}$ 
   } } ) 
 \nonumber \\
Q^8 \, (1) : && \qquad \qquad \qquad \qquad \qquad (0,0,0)
\nonumber
\eea
Here `perms' stands for all inequivalent permutations of the
three spins, and the number of
operators at each level is given in parenthesis. The scaling
dimension of $Q^n$ is $n/2$.
Tensoring these representations with those of a superconformal
primary gives the
spin content at each level of a long superconformal
representation.
 This   rombus   is truncated   in short
 multiplets because of  the appearance of null states. 
  
 Short ${\cal N}=4$ multiplets come in three  different series called $A_1$, $A_2$ and $B_1$  
 \bea\label{listr}
  A_1[j]_{1+j+R+R^\prime}^{(R;R^\prime)}\quad (j>0) \ , \qquad
  A_2[0]_{1+ R+R^\prime}^{(R;R^\prime)}\ , \quad {\rm and}\quad
  B_1[0]_{R+R^\prime}^{(R;R^\prime)}\ . 
 \eea
 The subscript $1,2$ indicates the level of the first null states. The $A_{1,2}$ multiplets appear in the
 decomposition of long multiplets at  unitarity threshold, so  their scaling dimension 
 can change continuously by recombination. The $B_1$ multiplets, on the other hand,  are separated   from the
 unitarity threshold by a gap.   
 Those with   $R<1$ or $R^\prime <1$ 
   don't  appear in the decomposition of any long representation. 
 Their dimension cannot therefore change without breaking
  ${\cal N} =4$ superconformal symmetry, and  is in this sense absolutely protected \cite{Cordova:2016emh}.
 
  \smallskip        
                       
 The most basic multiplets are   
  the  (16+16)-component multiplet of the  ${\cal N}=4$ 
  graviton in four dimensions, and the    (8+8)-component of the unbroken electic and magnetic gauge bosons. 
 The former includes the
graviton, six vectors and two scalars,
while the latter include a vector and six scalar fields.
  These fields   couple to the stress-tensor and   flavor-current multiplets  of the dual SCFT, 
 \bea\label{stress}
{\bf stress}:\quad
A_2[0]_{1}^{(0;0)} = [0]_{1}^{(0;0)} \oplus [0]_{2}^{(0;0)}
\oplus
 [1]_2^{(1;0) } \oplus [1]_2^{(0;1) } \oplus
[2]_3^{(0;0)}  \oplus {\rm fermions}
\eea
\bea\label{evec}
{\bf e-flavor}:\qquad B_1[0]_1^{(1; 0)} = [0]_1^{(1; 0)} \oplus
[1]_2^{ (0;0)}
\oplus [0]_2^{(0;1) }    \oplus {\rm fermions}
\eea
\bea\label{mvec}
{\bf m-flavor}:\qquad B_1[0]_1^{(0; 1)} = [0]_1^{(0; 1)} \oplus
[1]_2^{ (0;0)}
\oplus [0]_2^{(1;0) }  \oplus {\rm fermions}
\eea
 Notice the absence of a $\Delta =3$ scalar in the graviton multiplet, in agreement with   our earlier  
claim that the TsT mode  is not part of  the 4d supergravity spectrum.

The vector multiplets \eqref{evec} and \eqref{mvec} of  the
conserved  flavor-symmetry currents   belong to   absolutely-protected $B_1$   
     representations. This 
     implies that    
  flavor symmetries cannot break
continuously
   in   ${\cal N} =4$ superconformal theories. 
   Equivalently, 
gauge symmetries in the AdS$_4$ bulk cannot be broken \`a la
Higgs-Brout-Englert
without also breaking ${\cal N} =4$ supersymmetry, a conclusion  
    confirmed by the analysis of  Louis and Triendl
   \cite{Louis:2014gxa}.

     The 
stress-tensor multiplet, on the other hand, 
 belongs to the $A_2$ series of   representations  and does not enjoy the same kinematic protection. At unitarity
threshold one finds
     \bea
L[0]_1^{(0;0)} = A_2[0]_1^{(0;0)} \oplus B_1[0]_2^{(1;1)}\ . \eea
One may expect this recombination to occur when two decoupled
theories, with
separately-conserved energy-momentum tensors, are made to
interact very weakly so that only the total energy-momentum is exactly conserved.  
This still cannot happen continuously because ${\cal N} =4$
superconformal theories do not have {\it  any}
continuous moduli. It can however happen after a `small'
renormalization-group flow
     as   will be
discussed elsewhere.\,\footnote{C. Bachas and I. Lavdas, work in progress. 
A slightly massive graviton also
occurs in the Karch-Randall model
of locally-localized gravity  \cite{Karch:2000ct}.  The limit of 
vanishing graviton mass is, however, a decompactification limit 
  \cite{Bachas:2011xa}.} 
 
 \smallskip
 
 The next interesting set of representations are   
 $B_1[0]_{R }^{(R;0)}$ and $B_1[0]_{R^\prime
}^{(0; R^\prime)}$ for generic  $R$,  $R^\prime$.
 Their superconformal primaries are  always annihilated 
    by 4  out of the 8  Poincar\'e supercharges, 
so they are also     1/2 BPS.
  The superconformal primary with spin labels $(0,R ,0)$ for example is annihilated by the
supercharges $Q_\pm^{(+; \pm)}$.
These multiplets contain $16 R-8$ bosonic and as many fermionic
fields, all with   spins $\leq 1$,
        \vskip -4.2mm
     \bea \label{b1mult}
B_1[0]_{R }^{(R;0)} = [0]_{R }^{(R;0)} \oplus [1]_{R +1}^{
(R-1;0)}\oplus [0]_{R +1}^{ (R-1;1)}
   \oplus [0]_{R +2}^{(R-2;0)}  \oplus {\rm fermions}\ . 
   \eea
 These are the representations in which   open BPS strings on D5-brane stacks 
  transform. Note  that the spin-1 
component of the multiplet has S$^2$  angular momentum $ R-1$,  as   anticipated in the previous section.
   When $R=2$ it is  a vector of  SO$(3)_+$, precisely like   
   the   vector bosons corresponding to broken non-compact  electric 
 symmetries in the gauged-supergravity description,  SO$(3, m_+) \to$ SO$(3)_+\times G_+$
 (see section  \ref{gaugesugra}).  Similar statements hold of course    for the mirror
NS5-brane excitations which  
  have  Kaluza-Klein momenta   $(0;R^\prime)$.

        At the level of group theory it is possible to generate the representation  \eqref{b1mult}
by taking  tensor products of  the simplest ultrashort representation    
       \bea\label{ultrashort}
       B_1[0]_{1/2}^{(1/2;0)} = [0]_{1/2}^{(1/2;0)} \oplus  
       [\scalebox{0.9 }{ 
   ${1\over  2}\,$}
       ]_{1}^{(0;1/2)}\, .    
       \eea
  This is a
            free 
hypermultiplet  corresponding to a free superfield
 $H^a = q^a + \theta_\alpha^{a\dot a}\zeta_{\dot
a}^\alpha$
with $q^a$ a scalar doublet of ${\mathfrak{su}}(2)$, and
$\zeta_{\dot a}^\alpha$ a spinor doublet of
${\mathfrak{su}}(2)^\prime$.      
   The product of two  
hypermultiplets gives a conserved vector current
  multiplet
 \bea\label{twofree}
H^aH^b = q^{a} q^{b} + \theta_\alpha^{\{a\dot a} q^{b\}}
\zeta_{\dot a}^\alpha -
 \theta_\alpha^{ a\dot a}   \theta_\beta^{b \dot b}\,  
\zeta_{\dot a}^\alpha \zeta_{\dot b}^\beta\quad \to \quad
B_1[0]_{1}^{(1;0)}\ ,
\eea  
while,  more generally,  the product of $R$ identical hypermultiplets gives   $$
H^{a_1} \cdots H^{a_R}\quad  \to \quad   B_1[0]_{R }^{(R;0)}\ .   $$
The absence of spin 3/2 components in these representations
follows from the identity\,\footnote{The reader
may be amused to note that in the context of  the quark model    this same identity shows why   the  mysterious
$\Delta^{++}$ resonance requires the existence of three colors.}
    $$
{\rm Sym}_{abc}\, {\rm Sym}_{\alpha\beta\gamma}\, \Bigl[\, \theta_\alpha^{
a\dot a}
\theta_\beta^{ b\dot b}\theta_\gamma^{ c\dot c}\Bigr] = 0 \ .    $$
All this can be  repeated for  twisted
hypermultiplets $\tilde H^a = \tilde q^{\dot a} +
\theta_\alpha^{a\dot a}\tilde \zeta_{ a}^\alpha$ whose products
give the
mirror representations $B_1[0]_{R^\prime }^{(0;R^\prime)}$.   \smallskip

  The  $B_1$ multiplets with $R R^\prime \not=0$  contain  spin-3/2 components
 and  
  are 1/4 BPS (they continue 
 up to level $Q^6$).
 They  must be dual to  Kaluza-Klein modes of the 10d  
   gravitini, those  that  do not arrange themselves inside  spin-2  towers. 
  The lowest-lying  such multiplet,   
   $B_1[0]_{1 }^{(1/2;1/2)}$,    has as its  top   component a conserved supercurrent. 
It  is  present   in backgrounds with enhanced
supersymmetry,  but not in the solutions studied here 
which break   half of the ${\cal N}=8$ supersymmetries.

                    The  next entry in the list  \eqref{listr} are  the  
     representations $A_2[0]_{\Delta_0}^{(R; R^\prime)}$ with $R+R^\prime >0$ and
$\Delta_0 = 1+ R+R^\prime $.   These all contain spins up to and including 2 and are hence 
       Kaluza-Klein excitations of  the graviton. 
Like the stress-tensor multiplet, these multiplets  can  also
recombine into
      long representations at the unitarity threshold.  
  
      \smallskip    
             The last  set of   short super-conformal multiplets  is 
$A_1[j]_{\Delta_0}^{(R; R^\prime)}$ with $\Delta_0 = 1+
j+R+R^\prime $ and $j>0$. These
  contain spins higher than 2 and can only be dual to excited
string states,  or to multiparticle states.
     They  end at level 
$Q^4$ if $R=R^\prime=0$, at level $Q^6$ if $RR^\prime=0$, and at level
$Q^7$ otherwise. {With some abuse of terminology,  these special semi-short multiplets may be termed} accordingly  1/2, 
      1/4 or  1/8 BPS.  
       An example of such a representation is 
$A_1[1]_2^{(0;0)}$. It contains a conserved vector current at
the lowest level and
a conserved spin-3 current at level $Q^4$, and can become long
by eating
      $A_1 [\scalebox{0.9 }{ 
   ${1\over  2}\,$}
       ]_{5/2}^{(1/2;1/2)}$.  States  in the $A_1$ representations are interesting for the
       study of supersymmetric black holes, but they  are outside our  scope here.

For   later  reference we 
have collected  in  table \ref{tab:1} the short  ${\cal N}=4$
 superconformal multiplets and  corresponding 
single-particle fields in the AdS$_4$  backgrounds.


 \begin{table}
 \begin{center}
 \begin{tabular}{|c |c | c |  }
 \hline
  \, &\, &\,  \\  [-1.2ex]
  ${\cal N}=4 $ \bf Multiplet & \bf String mode  & \bf \,  Gauged SUGRA  \\
  [0.5ex]  \hline\hline \, &\, &\,  \\   [-1.2ex]
   $A_2[0]_1^{(0;0)}$  &   Graviton  & {\rm yes} \\    [0.5ex]
   \hline  \, &\, &\,  \\   [-1.2ex]
     $B_1[0]_1^{(1;0)}$  & D5 gauge bosons & {\rm yes} \\
   [0.5ex]  \hline  \, &\, &\,  \\   [-1.2ex]
  $B_1[0]_1^{(0;1)}$  & NS5 gauge bosons & {\rm yes} \\ 
    [0.5ex] \hline \, &\, &\,  \\   [-1.2ex]
   $B_1[0]_{R }^{(R>1;0)}$ &  \thead{  {Open F-strings}, \  
     $R \in   {1\over 2} \vert \ell_a - \ell_b\vert  + \mathbb{N}$  \\ Closed\  strings, \  $R\in \mathbb{N}$ }  
      & only $R=2$ \\ [0.5ex]
   \hline \, &\, &\,  \\  [-1.2ex]
$B_1[0]_{R^\prime }^{(0; R^\prime >1)}$ & \thead{ {Open D-strings}, \  
     $R^\prime  \in    {1\over 2} \vert \hat \ell_{\hat a} - \hat \ell_{\hat b} \vert + \mathbb{N} $
     \\ Closed\  strings,  \  $R^\prime \in \mathbb{N}$  }  & only $R^\prime=2$ \\ [0.5ex]
   \hline
    \, &\, &\,  \\  [-1.2ex]
  $B_1[0]_{R+R^\prime }^{(R\geq 1; R^\prime \geq 1)}$ &  
Kaluza Klein gravitini \  $({\scriptstyle R, R^\prime \in \mathbb{N}}) $ & {\rm no} \\   [0.5ex]
   \hline  
     \, &\, &\,  \\   [-1.2ex]
$A_2[0]_{1+ R+R^\prime }^{(R>0 ; R^\prime>0 )}$ & Kaluza Klein  gravitons  
 \  $({\scriptstyle R, R^\prime \in \mathbb{N}}) $ & no
 \\ [0.5ex]
   \hline 
      \, &\, &\,  \\  [-1.2ex]
$A_1[j>0]_{1+ j+R+R^\prime }^{(R ; R^\prime)}$ & Stringy
excitations & no  \\  [0.8ex]
     \hline  
  \end{tabular} \medskip\medskip\medskip 
\caption{\small The short ${\cal N} =4$ superconformal multiplets in the notation of ref.\,\cite{Cordova:2016emh}, 
possible  dual single-string excitations  in the Type IIB  solutions of   \cite{Assel:2011xz}\cite{Assel:2012cj},
and the  fate of these excitations after the gauged-supergravity truncation. 
 In the middle column    $\mathbb{N} = \{1,2,\cdots\}$ is the set of non-zero natural numbers. 
A yes/no entry in the third column indicates that the excitation  survives/does not survive
in gauged  ${\cal N}=4$ supergravity. 
Among the 
  $R=2$ or $R^\prime =2$    modes some
 may (but need not)  survive. }   \label{tab:1}
 \end{center}
 \end{table}
  

\subsection{Marginal   Deformations}       
Marginal  deformations are generated by scalar operators of
dimension $\Delta =3$.
To preserve maximal  supersymmetry they  must be top
components of
       ${\cal N} =4$ multiplets. 
       Inspection of all short multiplets  
shows that such  operators do not exist \cite{Cordova:2016xhm}, so   ${\cal N} =4$  SCFT$_3$  
        have  no {fully} superconformal moduli.
 We have seen  that the  same conclusion has  been   reached from gauged supergravity, 
 and  also from the explicit  Type IIB solutions.  The proof based on representations of 
 $\mathfrak{osp}(4\vert 4)$ settles definitely the issue.

  Actually, the                                 
   inspection of multiplets is tricky because some of
the 3d superfields have
top (or rather `dead end') components at intermediate levels
\cite{Cordova:2016xhm}.
       An example is  the scalar $[0]_2^{(0;0)}$
        in the stress-tensor multiplet  \eqref{stress} which
        is annihilated by all 
supercharges, and can be used to trigger a universal ${\cal N}
=4$ mass deformation\footnote{This
        deformation exists for any ${\cal N}\geq 4$. It  has
been discussed by many authors, especially in the context of the
ABJM theory. An incomplete list
of references is
\cite{Gomis:2008vc}\cite{Agarwal:2008pu}\cite{Kim:2010mr}\cite{Cheon:2011gv}.}.
       No such dead-end components arise however at  $\Delta =3$.  
The only other relevant ${\cal N} =4$ deformations reside in the
electric and magnetic
      flavor-current  multiplets  $B_1[0]_1^{(1;0)}$
      and $B_1[0]_1^{(0; 1 )}$ and 
       correspond 
to  triplets of flavor masses and   Fayet-Iliopoulos  terms    \cite{Intriligator:1996ex}\cite{Aharony:1997bx}. 
     
As shown in ref.\,\cite{Cordova:2016xhm} ${\cal N} =3$
also does not allow    fully superconformal moduli\footnote{{In principle  on can obtain
${\cal N} =3$ theories  from ${\cal N} =4$   by turning on a `quantised' super potential, 
that in ${\cal N} =2$ notation reads $W = k Tr(\Phi^2)$, and integrating out the massive modes. This produces Chern-Simons and supersymmetry related terms. In the brane setup it corresponds to combining the original 5-branes into (1,k) 5-branes.}}. 
The maximal supersymmetry that  allows them    is ${\cal N} =2$.
This is enough supersymmetry to protect some marginal operators
againts quantum corrections,
making the problem technically tractable. We focus henceforth on
${\cal N} =2$.

The ${\cal N} =2$ Poincar\'e supercharges are spinors $Q$ and
$\bar Q$ with R-charge,
respectively, $r=-1$ and $r=1$. The unitary multiplets are
two-sided: they are obtained
by imposing independent unitarity bounds and shortening
conditions for $Q$ and $\bar Q$.
All possible representations  are listed in pages  67-69 
 of ref.\,\cite{Cordova:2016emh}. The most relevant multiplets are the
conserved stress tensor,   vector current,
    and  `superpotential' multiplets \vskip -4mm
      \bea
    {\bf stress\ tensor }:\qquad
    A_1\bar A_1[1]_2^{(0)} = [1]_2^{(0)}\oplus
     [\scalebox{0.9 }{ 
   ${3\over  2}\,$}
       ]_{5/2}^{(\pm 1)} \oplus [2]_3^{(0)} \ , 
    \eea  
       \bea
   {\bf vector\ current}:\quad
     A_2\bar A_2[0]_1^{(0)} = [0]_1^{(0)}\oplus 
      [\scalebox{0.9 }{ 
   ${1\over  2}\,$}
       ]_{3/2}^{(\pm 1)}
     \oplus [0]_2^{(0)} \oplus [1]_2^{(0)}\ , 
    \eea
      \bea
     {\bf superpotential }:\quad
L\bar B_1[0]_r^{(r>0)} = [0]_r^{(r)}\oplus  [\scalebox{0.9 }{   ${1\over  2}\,$}
       ]_{r+{1\over 2}}^{(r-1)} \oplus [0]_{r+1}^{(r-2)}\ . 
    \eea
These couple respectively 
to the   graviton multiplet, the vector boson multiplets,  and  the   
 hypermultiplets of  4d ${\cal N} =2$  supergravity. 
   The  first two representations are self-conjugate, 
while the third is paired with  the antichiral  multiplet    $\ B_1\bar L[0]_{-r}^{(r< 0)} $. 
 
    \smallskip

{
                                       
The local structure of  the superconformal manifold of   
  3d ${\cal N} =2$ CFTs 
is essentially the same as for  4d  ${\cal N}=1$ theories, 
and   is well understood \cite{Kol:2002zt,Benvenuti:2005wi}.
   Firstly, candidate moduli  only exist in the
 marginal superpotential  multiplet 
$
L\bar B_1[0]_2^{(2)}$ and its conjugate \cite{Cordova:2016xhm}.
They  may, however, fail to be exactly marginal because these
 multiplets are not absolutely
protected and can recombine in long multiplets at the unitarity
threshold. The power of the superconformal algebra is that it points to a
unique culprit:
the only possible  recombination is  with a vector current \bea\label{higgs}
L\bar B_1[0]_2^{(2)} \oplus B_1\bar L[0]_{-2}^{(2)} \oplus
A_2\bar A_2[0]_1^{(0)}
   \  \to\  L\bar L[0]_1^{(0)}\ .
\eea
This is the only mechanism by which a marginal operator ${\cal L}$ can
become marginally irrelevant in 3d  ${\cal N} =2$ 
{providing   the `longitudinal' component of a previously conserved current 
${\cal J}_\mu$\,\footnote{As shown in \cite{Bianchi:2003wx}-\cite{Bianchi:2006ti}, this generalises to generic spin $s$ currents in (super)conformal theories, \vskip -3mm
$$
\partial{\cal J}_{(s,\Delta=s+D-2)} = 0 \quad \rightarrow \quad \partial {\cal J}_{(s,\Delta=s+D-2+\gamma)} = {\cal L}_{(s-1,\Delta=s+D-1+\gamma)}  \ ,
$$
provided that the
 required `Goldstone' or St\"uckelberg modes ${\cal L}_{(s-1,\Delta=s+D-1+\gamma)}$ exist  in the spectrum. 
} 
 \bea\label{longitudinal}
\partial_\mu {\cal J}^\mu = 0 \quad \rightarrow \quad \partial_\mu {\cal J}^\mu = {\cal L} \ .
\eea
 \noindent In the end, as shown using only superconformal perturbation theory
in  \cite{Green:2010da}, the superconformal manifold  ${\cal M}_c$ is the K\"ahler quotient 
of the
space $\{\lambda_i\}$
of complex marginal couplings by the complexified global (flavor)
symmetry group $G$,  
\bea\label{quotient}
{\cal M}_c =  \{\lambda_i \vert D^a=0\}/G = \{\lambda_i\}/G^{\mathbb{C}}\ . 
\eea
Here  $D^a=0$ is the  moment-map  condition
   \bea\label{quotient1}
D^a =  \lambda^i T^a_{i\bar j} \bar\lambda^{\bar j} +  O(\lambda^3)\  
=\ 0
   \eea 
with   $T^a$  the 
generators of the global group
in the representation of the marginal couplings.  This condition follows directly from conformal
perturbation theory, while the extra  quotient  by $G$ is just the identification of theories
obtained by $G$-transformation of the deforming operator \cite{Kol:2002zt,Benvenuti:2005wi}.

   The story is more  familiar  in the context ${\cal N}=1$ SCFTs in four dimensions,
   following the  pioneering work  of Leigh and Strassler  \cite{Leigh:1995ep} and  further
   explored in  \cite{Kol:2002zt,Benvenuti:2005wi,Aharony:2002hx,Asnin:2009xx}.
From the perspective of the dual supergravity the recombination
\eqref{higgs} is the familiar `Higgsing' of a gauge symmetry which is allowed
by ${\cal N} =2$, but not by ${\cal N} =4$ 
supersymmetry.\,\footnote{{${\cal N} =4$ supersymmetry is however compatible with a pantagruelic Higgs mechanism, termed {\it La Grande
Bouffe} in \cite{Bianchi:2003wx}-\cite{Bianchi:2006ti}, whereby higher spin currents are violated by the interactions and acquire anomalous dimensions $\gamma$ in the boundary (S)CFT, holographic dual to mass-shifts in the bulk AdS (super)gravity. The prototypical example is the long Konishi super-multiplet, originally studied in e.g. 
\cite{Andrianopoli:1998ut}-\cite{Bianchi:1999ge}.}} 

When  the CFT  has no flavor symmetries
the supergravity has no vector multiplets. Marginal deformations
are dual to hypermultiplets 
and since ${\cal N} =2$ forbids a superpotential  they are,  in this case, 
exactly marginal. This agrees with the fact that the  quotient \eqref{quotient}  is in this case
the  trivial quotient. When the supergravity  couples  to 
vector multiplets the analysis of the vacua is   more involved, see for example  \cite{deAlwis:2013jaa}.
       \smallskip

 }     
            
       
\subsection{ $\boldsymbol{{\cal N}=4 \ \to \  {\cal N}= 2}$ }

We are   interested in marginal deformations of  ${\cal N}=4$ theories preserving ${\cal N} =2$. 
Such  deformations only exist   in
   the   ${\cal N} =2$ superpotential multiplet 
 \bea\label{superpotential} 
L\bar B_1[0]_2^{(2)} = [0]_2^{(2)} \oplus  [\scalebox{0.9 }{ 
   ${1\over  2}\,$}
       ]_{5/2}^{(1)} \oplus  [0]_3^{(0)} \  
 \eea
and its conjugate $B_1\bar L$. 
Since our starting point is ${\cal N} =4$ supersymmetric, 
 we must search for  
${\cal N} =4$ multiplets that contain such  marginal ${\cal N} =2$ operators.
  The  ${\cal N} =2$ subalgebra 
  has a unique embedding in ${\cal N} =4$,   and commutes with a $u(1)_F$ that acts
  as an ``accidental flavor symmetry'', 
\bea\label{pick}
 \mathfrak{osp}(2\vert 4)\oplus \mathfrak{u(1)}_F  \subset \mathfrak{osp}(4\vert 4)  
\eea
In  an appropriate  basis   the  ${\cal N} =2$ R-symmetry  is
generated by $\mathfrak{J}_3 +
\mathfrak{J}_3^\prime$, where $\mathfrak{J}_3$ and
$\mathfrak{J}_3^\prime$ are the
 canonically-nomalized Cartan 
generators of  $\mathfrak{su}(2)\oplus \mathfrak{su}(2)^\prime$,  while   the ``accidental'' 
commuting $\mathfrak{u}(1)_F$ is  
 generated by $\mathfrak{J}_3 - \mathfrak{J}_3^\prime$. We are interested in the 
 decomposition of ${\cal N} =4$
 multiplets  under this embedding.
 
}

 \smallskip

 The relevant  ${\cal N} =4$ representations must  
 contain
 a  $\Delta=2$ scalar component, which is the lowest component of \eqref{superpotential}.  
  This  gives the following candidate   list
\bea\label{list}
B_1[0]_{R+R^\prime}^{(R;R^\prime)}  \quad  {\rm with}  \ \ R+R^\prime = 1,2  \ , 
\qquad {\rm or} \qquad   A_2[0]_{R+R^\prime+1}^{(R;R^\prime)}  \quad  {\rm with}  \ \ R+R^\prime = 0,  1\ . 
\eea
  The  lowest  entries of the   list   are the 
familar stress-tensor and vector-current
multiplets, eqs.\,\eqref{stress} - \eqref{mvec},  whose    ${\cal N} =2$ decomposition 
     reads:\vskip -4mm
 \bea\label{4to2gr}
  A_2[0]_{1}^{(0;0)} \  = \ 
  \underbrace{A_2\bar A_2[0]_1^{(0){\color{blue} (0)}} 
  }_{\rm vector  \ current}
 \  \oplus \ 
 \underbrace{
A_1\bar A_1 [\scalebox{0.9 }{ 
   ${1\over  2}\,$}
       ]_{3/2}^{(0){\color{blue} (1)}}\oplus 
A_1\bar A_1 [\scalebox{0.9 }{ 
   ${1\over  2}\,$}
       ]_{3/2}^{(0){\color{blue} (-1)}}
}_{\rm supercurrents}
 \  \oplus \ 
\underbrace{
A_1\bar A_1[1]_{2}^{(0){\color{blue} (0)}}
}_{\rm stress \ tensor}
 \ , 
 \eea
 \bea\label{4to2e}
  B_1[0]^{(1; 0)}=  \ 
   \underbrace{ A_2\bar A_2[0]_1^{(0){\color{blue} (0)}}
   }_{\rm vector  \ current} 
  \  \oplus  \ 
   \underbrace{L\bar B_1  [0]_1^{(1){\color{blue} (1)}}
  \oplus 
  B_1\bar L [0]_1^{(-1){\color{blue} (-1)}}
  }_{\rm chiral  + antichiral} 
    \ , 
\eea
\bea\label{4to2m}
B_1[0]^{(0; 1)}= \ A_2\bar A_2[0]_1^{(0){\color{blue} (0)}}
\oplus L\bar B_1 [0]_1^{(1){\color{blue} (-1)}}
  \oplus B_1\bar L [0]_1^{(-1){\color{blue} (1)}} . 
 \eea
These are the familiar decompositions of the ${\cal N} =4$
 graviton and
 vector multiplets  in terms of ${\cal N} =2$ multiplets. 
The conserved current in the graviton multiplet  is that of the
`accidental'
$\mathfrak{u}(1)_F$  symmetry. We indicated  the $\mathfrak{u}(1)_F$   charge 
of representations   in blue fonts 
next to the ${\cal N}=2$ R-symmetry charge.

None of these multiplets  contains  the sought-for
marginal superpotential   \eqref{superpotential}.   
One  can also rule out   
the  representations $A_2[0]_2^{(R;R^\prime)}$ 
with $R+R^\prime =1$ from the list, because 
the $\Delta=2$ scalars in this representation have
${\cal N}=2$  R-charge $r= 0, \pm 1$, 
but not $r=2$ as required.  This leaves  then  the representations 
  $B_1[0]_2^{(R;R^\prime)}$  with $R+R^\prime = 2$ as the only ones containing marginal
  ${\cal N}=2$ superconformal deformations. 
 \smallskip

The decomposition of these   representations in terms of
${\cal N} =2$ multiplets is as
follows\footnote{In comparing with the tables   in
ref.\,\cite{Cordova:2016emh} the reader should
    be warned 
that  these  are only valid for generic
values of the spins. For small $R, R^\prime$ some of the   states  are actually  missing.}
    [we only exhibit one member of each  mirror pair;  
      the 
     c.c.  of $X\bar Y[j]_\Delta^{( r){\color{blue} (f)}}$  is 
 $Y\bar X[j]_\Delta^{( -r){\color{blue} (-f)}}$ 
    ]:
    \vskip -1mm
  \bea\label{1}
B_1[0]^{(2;0)}_2 \ = \ L\bar L[0]^{(0){\color{blue} (0)}}_2
\oplus
 \Bigl[      L \bar A_2 [0]_2^{(  1){\color{blue} (1)}} 
     \ \oplus\   
   \, 
   \boxed{ L\bar B_1[0]_2^{(2){\color{blue} (2)}}   } 
    \, \oplus 
    c.c. \Bigr]\ , 
\eea
 \bea\label{3}   
B_1[0]^{(1;1)}_2 \ = \ && \hskip -0.6cm L\bar
L[0]^{(0){\color{blue} (0)}}_2
        \oplus   
  \Bigl[   L\bar L[0]^{(0){\color{blue} (  2)} }_2 \oplus  
   L \bar A_2 [0]_2^{(  1){\color{blue} (  1)}} \oplus
   L \bar A_2 [0]_2^{(  1){\color{blue} (-1)}} 
    \oplus  
    \boxed{  L\bar B_1[0]_2^{(2){\color{blue} (0)}   } }
    \oplus
    \,  c.c. \Bigr] \nonumber 
 \\  \,    \nonumber \\
&& \hskip -0.4cm \oplus \,\, \Bigl[ L\bar L [\scalebox{0.9 }{
   ${1\over  2}\,$}
       ]^{(0){\color{blue} (  1)}}_{5/2}   \oplus 
          L\bar A_1  [\scalebox{0.9 }{ 
   ${1\over  2}\,$}
       ]_{5/2}^{(1){\color{blue} (0)}}  \oplus\,  
   c.c.\ 
   \Bigr]  \oplus\,   L\bar L[0]^{(0){\color{blue} (0)}}_3\ , 
  \eea
  \bea\label{2}
B_1[0]^{(3/2; 1/2)}_2 \ = \ && \hskip -0.6cm \Bigl[ L\bar
L[0]^{(0){\color{blue} (1)}}_2
        \oplus  
   L\bar A_2[0]^{(1){\color{blue} (  2)} }_2 \oplus  
   L \bar A_2 [0]_2^{(  1){\color{blue} (0)}} 
   \, \oplus   \, 
   \boxed{ L\bar B_1[0]_2^{(2){\color{blue} (1)}   }} \oplus
    \,  c.c. \Bigr]  \nonumber 
 \\  \,    \nonumber \\
&& \hskip -0.4cm \oplus \,\, L\bar L[\scalebox{0.9 }{ 
   ${1\over  2}\,$}]^{(0){\color{blue} (
0)}}_{5/2} \oplus
\Bigl[   L\bar A_2 [0]_{3}^{(2){\color{blue} (1)}}  \oplus\,   c.c.\ 
   \Bigr] \ .  
\eea  
 \vskip 2mm
 \noindent   
All of them contain the (boxed) marginal superpotential $L\bar
B_1$, so we should look for these   multiplets in the original ${\cal N} =4$ theory.

The breaking ${\cal N}=4 \to {\cal N}=2$
     requires that two spin-1/2 Goldstini   be eaten
by the two spin-3/2 gravitini
which  acquire  a mass. The corresponding   recombination  reads    
      \bea
     A_1\bar A_1 [\scalebox{0.9 }{ 
   ${1\over  2}\,$}
       ]_{3/2}^{(0) }  
     \oplus L\bar A_2[0]_2^{(1) } \oplus c.c. \ \ \to \ \ 
 L\bar L [\scalebox{0.9 }{ 
   ${1\over  2}\,$}
       ]_{3/2}^{(0)}    \ ,    
          \eea
          where $L\bar A_2[0]_2^{(1) } $ is the  ${\cal N}=2$ Goldstino multiplet. 
    All of the above muliplets contain, in addition to the deforming superpotential,       
   a candidate Goldstino as required for consistency. 
   Note also that  the self-conjugate representation \eqref{3}   respects  the accidental
$\mathfrak{u}(1)_F$ symmetry, while 
among all     flat directions in  other multiplets  
 a  linear combination will be   lifted by the
        $\mathfrak{u}(1)_F$  moment-map condition.

  To summarize, we have identified  the three  ${\cal N}=4$ superconformal multiplets that
     contain candidate  ${\cal N}=2$  moduli. 
      From   table 
 \ref{tab:1} we see    that  the first can arise from either  open or closed strings, and may  survive 
 the  gauged-supergravity truncation,
 the second can come from Kaluza-Klein gravitini, while the third  violates
 our angular-momentum selection rule of  section  \ref{3.1} and can only be  an exotic
  multi-particle state.  We will now go to the  CFT side of the correspondence, where the
    first kind  of deformations        
         will be identified with standard  superpotential deformations   on  the
         Higgs branch of the electric-quiver gauge theory (or its mirror dual),  
         while the others   
        involve mixed-branch operators.


  
 \section{ $\boldsymbol{{\cal N}=4}$   quiver gauge  theories }
 \label{QGT}

  The  evidence that the
   AdS$_4$ solutions of  \cite{Assel:2011xz}\cite{Assel:2012cj}
    are   holographic duals to the    `good' 
   ${\cal N}=4$  quiver gauge  theories of
      \cite{Gaiotto:2008ak} is that  their symmetries match, and that 
     the 
    condition  \eqref{order}    is automatically obeyed
    on the string-theory side. In this section we will find more  evidence
    for this  correspondence. 
      To make the paper self-contained, we start  by  recalling  some simple  
        facts about  ${\cal N}=4$  quiver  theories.

\subsection{Generalities}
\label{lqgt}

     The     fields  of  ${\cal N}=4$ gauge  theories  are   vector multiplets and hypermultiplets,
     and the  group of R symmetries is usually denoted 
     SU(2)$_H\times$SU(2)$_C$
     (for Higgs and Coulomb).\footnote{This is the same as  the  group SO(3)$_+\times$SO(3)$_-$ of earlier sections.  } 
     A  vector multiplet  contains  a vector, $A_\mu$,  and three  scalars $\phi^{1,2,3}$
     transforming  as a   vector   of  SU(2)$_C$, while a hypermultiplet contains two complex
     scalars,  $H= (q^+, q^-)$,  which  transform as a doublet of SU(2)$_H$.

 \begin{figure}[t!]
\centering 
 \vskip -0.9 cm
\includegraphics[width=.83\textwidth,scale=0.60,clip=true]{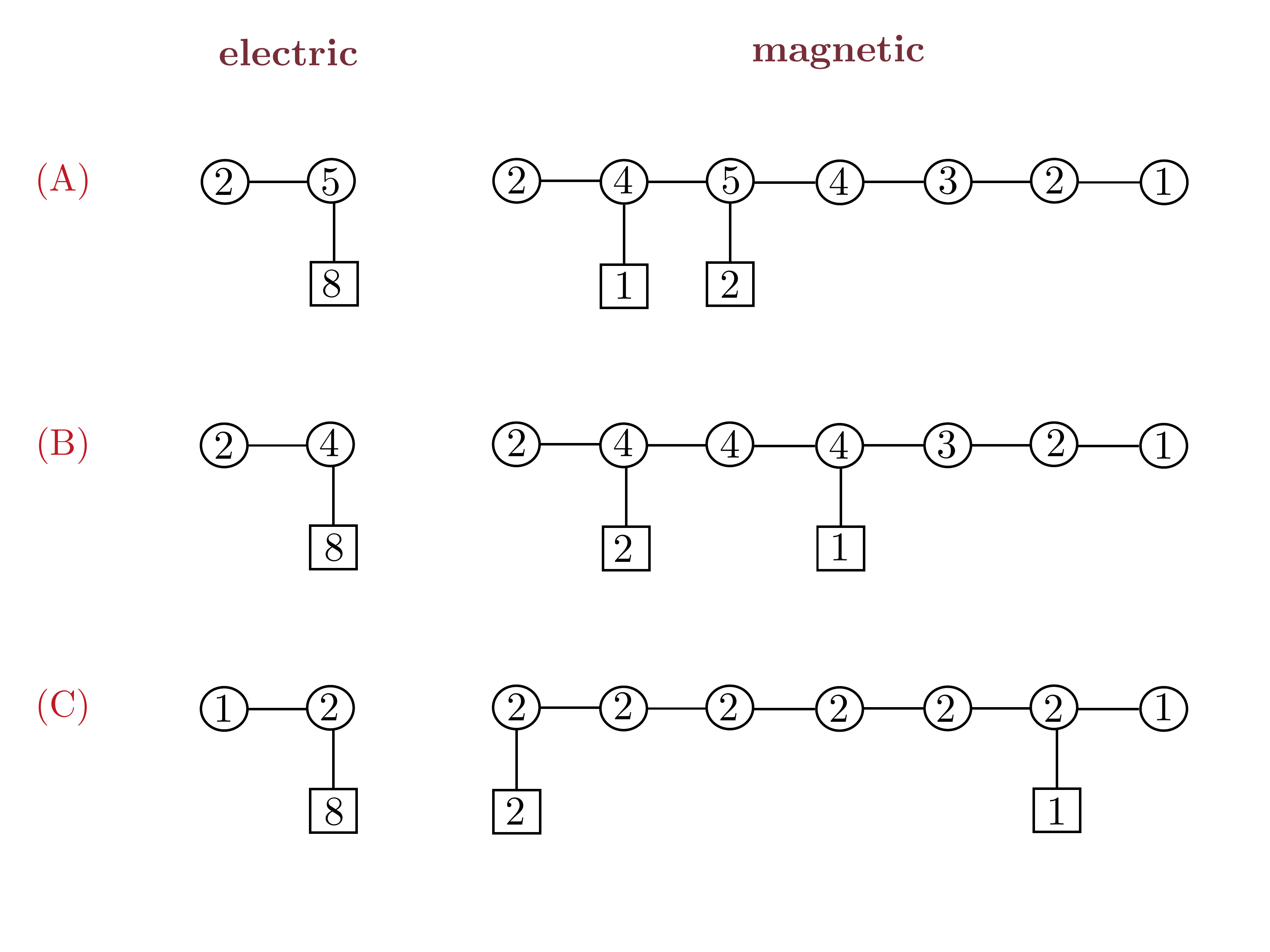}
  \vskip -6mm
   \caption{\small The electric and magnetic quivers corresponding to  the    theories $(A,B,C)$   
   of section \ref{gravside}\,. The infrared limits  of these theories have the same flavor symmetry
    but different   operator content. }
\label{fig:2} 
 \end{figure}

       The vector fields
     are in  the adjoint representation of the gauge group which is a product  of unitary factors, 
     $\prod_i U(n_i)$,
     while the hypermultiplets transform either in  fundamental  or in bifundamental representations.
    The precise representation content is described by a linear quiver   like  those depicted in
       figure \ref{fig:2}\,. A    circular node
      \circled{$n_i$} stands for a gauge-group  factor $U(n_i)$, a square node $\boxed {N_a}$ for 
      $N_a$ hypermultiplets in the fundamental of the associated gauge-group factor, and 
      a horizontal link for a hypermultiplet in the bifundamental 
      of the adjacent circular nodes. 
     The   ${\cal N}=4$ Lagrangian contains the Yang-Mills term but no Chern-Simons 
     terms,\footnote{Chern-Simons terms are necessary when there exist three different types
     of 5-brane, in which case the maximal supersymmetry is ${\cal N}=3$
     \cite{Kitao:1998mf}. The ${\cal N}=4$ theories of interest here  have equivalent Chern-Simons
     realizations   \cite{Assel:2017eun}, which  we   will not need   in this paper. 
     }
     and it is  completely
     fixed  by the quiver  data modulo the   gauge couplings  which are  free parameters  with  dimension  
     $[{\rm mass}]^{1/2}$.

      The relation of the quiver to the data $\{N_a, \ell_a\}$ and  $\{\hat N_{\hat a}, \hat \ell_{\hat a}\}$ of
     the supergravity solutions  follows from the   engineering  of the 
     gauge theory  on flat-space branes  \cite{Hanany:1996ie}. 
     The electric quiver is realized 
      by 
     D3-branes suspended between NS5-branes and intersecting D5-branes, while for the 
     magnetic quiver the roles of NS5 and D5 are exchanged. 
     It follows   from these constructions that the 
      electric quiver  has $N_a$ fundamental 
      hypermultiplets at the $i=\ell_a$ gauge node (counting from right to
      left),  while the magnetic quiver has $\hat N_{\hat a}$ fundamental 
      hypermultiplets at the $\hat i= \hat\ell_{\hat a}$ gauge  node (counting from left  to
      right).  The number of gauge nodes in   the electric
   quiver  is  the total number $\hat k= \sum_{\hat a} \hat N_{\hat a}$  of NS5-branes  minus one.    
        Reading   the  complete  NS5-brane  data
      $\{\hat N_{\hat a}, \hat \ell_{\hat a}\}$   from  the electric quiver,
        is  possible but slightly  more involved
      (see e.g.   ref.\,\cite{Assel:2011xz}).  By moving around the 5-branes one obtains   the dual 
      magnetic quiver with a  total number of gauge nodes $\sum_a N_a -1:= k-1$.

       The   condition  \eqref{order} for a `good theory'  ensures that the number of  hypermultiplets
        suffices to completely Higgs the gauge symmetry. 
       The  vacuum manifold    thus contains  a pure Higgs branch, along which only the 
       hypermultiplet v.e.vs are  non-zero,  and a pure Coulomb branch which is  isomorphic to
       the pure Higgs branch of the mirror  quiver.  These branches can be viewed as 
       complex varieties  described by  chiral rings of holomorphic functions. 
           In the language  of ${\cal N}=2$ supersymmetry, 
     vectors  decompose   into  pairs $(V, \Phi)$ of  vectors and chiral multiplets  in the adjoint
       representation of the gauge group, while  hypermultiplets  contain two chiral fields
       $(q, \tilde q)$ in complex-conjugate representations.\,\footnote{Compared to
       our previous  notation,  $q := q^+$ is  the upper component of the $SU(2)_H$ 
doublet,     and $\tilde q := (q^-)^\star$ is  the complex conjugate of the   lower component.
       } 
       The Yang-Mills Lagrangian 
       includes a superpotential term  for each $U(n_i)$  factor, 
       $W =  \sum_{\alpha} q_\alpha^T   \Phi \tilde q_\alpha$ where
       $\Phi$ is the adjoint chiral field and
           the sum runs over all 
       chiral multiplets   in the (anti)fundamental representation of $U(n_i)$. 
   Since on  the Higgs branch $\langle \Phi \rangle = 0$, 
     the   non-trivial F-flatness conditions are  the $n_i^2$ equations 
        $\sum_{\alpha} q_\alpha   \tilde q_\alpha^T = 0 $ for  each gauge node.  The chiral ring
      consists of all gauge-invariant  combinations of  $q$s and $\tilde q$s   modulo
        these  F-flatness conditions.


  \subsection{Chiral-ring operators  as open strings}
 
          The chiral-ring operators on the  Higgs branch are singlets of  $SU(2)_C$. Their dimension
          is equal to their ${\cal N}=2$ R-symmetry charge which is half  the number of chiral fields in a product.
          It is easy to see that all such  operators are  highest-weights in  short representations
          $B_1[0]_R^{(R;0)}$ of the ${\cal N}=4$ algebra,   where $\Delta = R$ is the $SU(2)_H$
             spin.  These representations   
         survive the RG flow to  the infrared SCFT.

\begin{figure}[t!]
\centering 
 \vskip -1.8 cm
\includegraphics[width=.85\textwidth,scale=0.70,clip=true]{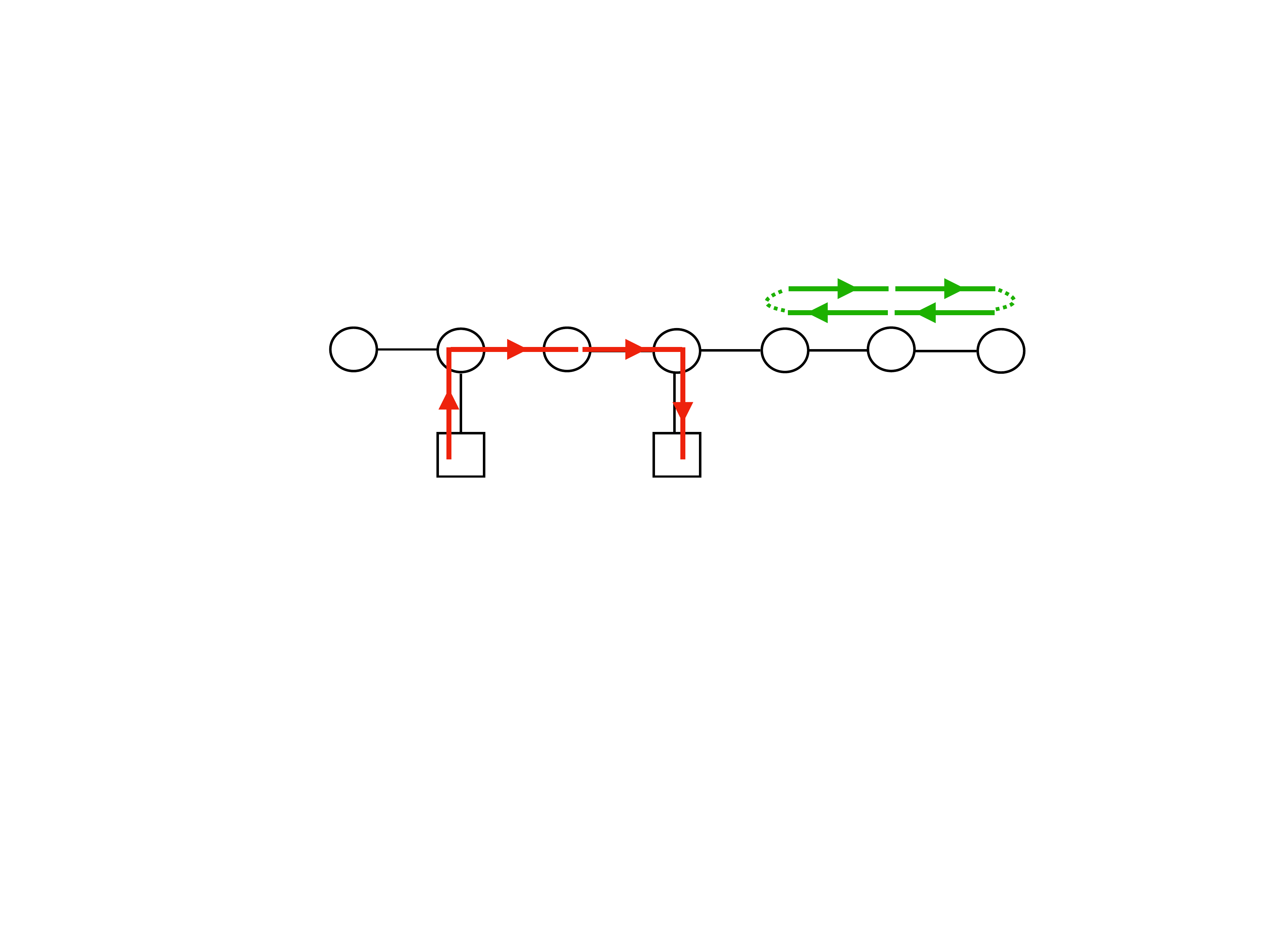}
  \vskip -35mm
   \caption{\small  Two chiral operators  on the Higgs branch of the magnetic quiver of theory B. 
   The open string operator (in red) is in the bifundamental of the
   flavor group $U(2)\times U(1)$, while the closed string operator (in green) is a flavor singlet. 
   Both contain  marginal superpotential deformations since they have  length=4,  
   and hence belong to     $B_1[0]_2^{(2;0)}$ multiplets.}
\label{fig:3} 
 \end{figure}

        One can  picture these operators as oriented strings with  string-bits being the links of
        the quiver diagram, as in  figure \ref{fig:3}. {The usefulness of this  perspective
        was demonstrated by Assel   \cite{Assel:2017hck} who mapped chiral and monopole operators to
       strings in the flat-brane setup. Here we will map them   to strings in the near-horizon geometry.}
        The string orientation indicates whether one picks  the $q$ or $\tilde q$ chiral field in  a
        hypermultiplet:  the $q$ for  up- or left-pointing string bits, and the 
         $\tilde q$  for   down-  or right-pointing string bits.  
              Gauge-group indices are automatically summed while flavor indices are
        free, so strings  can be  either closed,  or have  endpoints on the square nodes
        which we identify with the stacks of  D5-branes.  
       The same holds  for the dual magnetic quiver,  where the square nodes are 
        NS5-branes and the relevant  ${\cal N}=4$ multiplets are $B_1[0]_{R^\prime}^{(0: R^\prime)}$.
        Clearly the dimension $\Delta$ is half the total  length of the string.

\begin{figure}[b!]
\centering 
 \vskip -0.6 cm
\includegraphics[width=.80\textwidth,scale=0.70,clip=true]{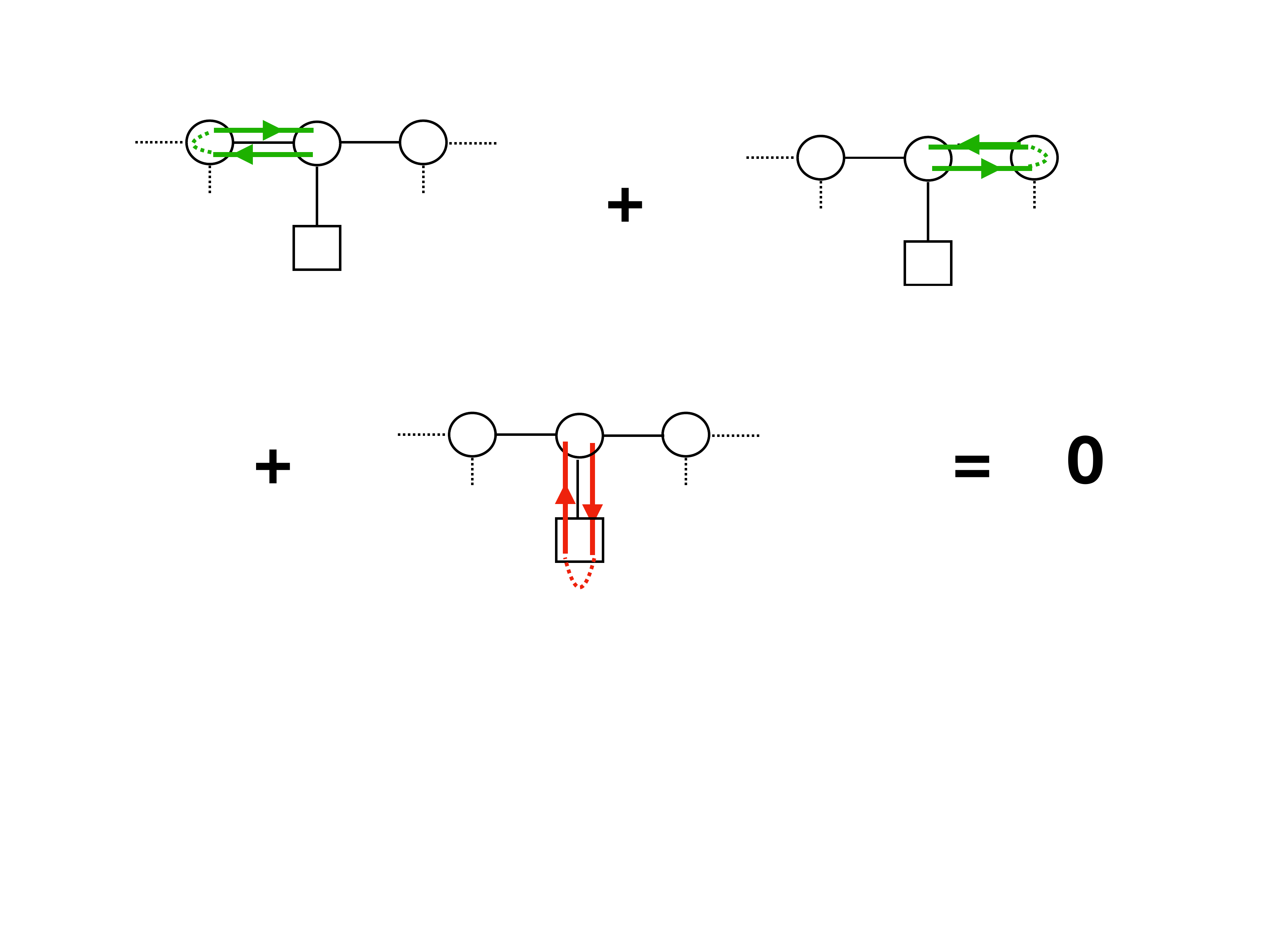}
  \vskip -35mm
   \caption{\small  Graphical representation of the
    F-flatness conditions on the Higgs branch,  as linear relations among
   cut-open string segments. 
   The dotted red semicircle  in the third term stands  for a summation over  the free  flavor indices
   of the open strings.  }
\label{fig:4} 
 \end{figure}

   To make the notation lighter it is convenient to  label the D5-brane stacks by their  
    linking number,   indicating  the circular node to which they attach. The quiver data
    is then specified by the   set of $\hat k-1$ non-negative integers $S= \{ N_\ell\}$,  some of which
    can  be zero, and by the set of corresponding gauge-group ranks $s= \{n_\ell\}$ .  
    For example  the   magnetic quiver of theory A  in figure \ref{fig:2}  has 
    $S = \{ 0, 1, 2, 0, 0 , 0, 0\}$ and $s=\{2, 4, 5, 4, 3, 2, 1\}$.  We also consider the $q$ and $\tilde q$ as
    matrices, so that the superpotential
    at the  $\ell^{th}$ circular node 
    reads   
   \bea W = {\rm tr}(\Phi_\ell \, \tilde q_{\ell,\square}  \,q_{\square,\ell} +  \Phi_\ell\,  \tilde q_{\ell, \ell+1}\, q_{\ell +1 , \ell }
     +  \Phi_\ell \,  q_{\ell, \ell-1 }\, \tilde q_{\ell-1, \ell })
   \eea
  where the subsripts here indicate the link,  with   gauge and flavor indices suppressed (the label `$\square,\ell$'  
  denotes the vertical link).   
    The corresponding F-term condition can
          be drawn  as   a linear relation   between   strings  cut-open  at  the    circular node,
          as shown   in figure  \ref{fig:4}.  Notice that this relation identifies  
          flavor-singlet combinations of open strings with   closed strings.

  \medskip

   Having set the  notation, let us   enumerate some  chiral operators by increasing length:

   \begin{itemize}
   
   \item     {$\mathbf{\Delta = 1}$}.  These   operators belong to  conserved-current multiplets  of ${\cal N}=4$. 
   There exist length-2    open strings in the adjoint representation of  {each $U(N_\ell)$
   flavor-group factor, }
    and one 
   length-2  closed string 
   for each  horizontal node of the quiver, i.e. in total  $\hat k-2$ horizonatl nodes.   
   These strings  are subject to    $\hat k-1$  F-term 
    conditions, one for each   
      $tr \Phi_\ell$.  The number of independent operators is therefore 
      precisely the dimension of  the flavor group $\bigl( \prod_\ell U(n_\ell)\bigr)/U(1)$
      as expected.  
   Note  that the overall $U(1)$ which acts trivially on the  fields   consistently decouples.

   \item    {$\mathbf{\Delta = 3/2}$}.  There are no closed strings of length 3, in accordance with
   the   rule on the gravity side,  see  section  \ref{excitations}, 
       that  spin-0 closed strings must have integer S$^2$  angular momenta.
   Open strings of length 3 exist for every  neighbouring  pair  of  square nodes,  and they are linearly independent.
 They   
    transform in the bifundamental representation of $U(N_\ell)\times U(N_{\ell+1})$.  
      Of the six quivers of figure \ref{fig:2} only the  magnetic quiver of  theory  A
   has such chiral operators. 
   \vskip -5mm
   
   \item  {$ \mathbf{\Delta = 2}$ }.  There are several possibilities for strings of length 4.
   The corresponding chiral operators belong to the short multiplets $B_1[0]_2^{(2;0)}$ (or  $B_1[0]_2^{(0;2)}$ for the Coulomb branch or, alternatively, the Higgs branch of the dual quiver) which  contain  marginal ${\cal N}=2$ superpotential deformations. 
   \, \hfil \break
The open strings transform either in the second symmetric product of the flavor group, or in bifundamental representations of next-to-nearest-neighbor group factors $U(N_\ell)\times U(N_{\ell+2})$, when such exist. 
   Of the six quivers of figure \ref{fig:2} only the  magnetic quiver of  theory  B 
   has such chiral operators.   
   The operators in the second symmetric product of the adjoint representation 
  are bound states of two open strings. We will argue shortly that this completes
   the list of independent $\Delta =2$ chiral operators. 
    
   \end{itemize}
 
 \vskip -3mm
       \noindent    One can now see the emerging general pattern.   At  level  $\Delta = R$ there exist single-string chiral
          operators that are either closed strings,  or open  strings 
          in the bifundamental of $U(N_\ell)\times U(N_{\ell^\prime})$.  They are subject
          to the following selection rules:
        \bea
         R = 1+n  \quad {\rm for\ closed\ strings}, \qquad 
         R = {1\over 2} \vert \ell - \ell^\prime\vert + n   \quad {\rm for\ open \ strings}
          \eea  
        where $n=1, 2 \cdots$.  These  are precisely the rules 
         derived  from  Type IIB string theory   in section \ref{excitations}\,,  
         in perfect  agreement  with  the conjectured holographic duality.  {They have been   also 
         obtained as scaling dimensions of monopole operators on the Coulomb   branch of the
          magnetic quiver    in agreement with mirror symmetry \cite{Borokhov:2002cg}.}

          
               For linear quivers, it is actually possible to choose a basis   in which all chiral operators are multiparticle 
               bound states of open strings.  This follows from the F-term condition in figure \ref{fig:4}
               \bea\label{Fold}
               -  \tilde q_{\ell, \ell+1}\, q_{\ell+1, \ell }\,=\,      q_{\ell, \ell - 1}\, \tilde q_{\ell-1, \ell} + \tilde q_{\ell,\square}  \,q_{\square,\ell}\ , 
               \eea
          which can be used to ``fold and slide''  
          any  closed string along the horizontal part of the quiver.  If the second term on the right-hand-side were
          absent, the closed string would eventually hit the quiver boundary and annihilate.  Because  of the second term,
          the process is accompanied by the ``emission''  of open strings, {\em qed}. 
                                  
             This argument does not   work for circular quivers,  which have no boundary and which can support 
             irreducible closed winding strings.


\subsection{Counting techniques}
\label{4.3}

     Although the logic is clear,  the actual 
          counting of  $\Delta >  2$ operators on the pure Higgs or pure  Coulomb 
          branch  can become  quickly  cumbersome.   In this section we mention 
          some elegant techniques developped by both physicists and mathematicians.

  The moduli spaces of the family of ${\cal N}=4$   quivers  considered here   have a 
 description in terms of nilpotent orbits and Slodowy slices, 
 see  \cite{Cabrera:2017njm} and references therein. 
  Let us review the basic points. 
 For a given partition $\rho$ of $N$ 
  we denote by $\overline{\cal O}_\rho$ the closure of the nilpotent orbit associated to this partition.
  The orbit  consists of all $N\times N$ nilpotent matrices  whose Jordan normal form has Jordan 
  blocks of sizes given by the partition, and the  closure includes the orbits of all  smaller partitions as well. 
  We also denote  by ${\cal S}_\rho$ the Slodowy slice associated to this partition, 
   namely  the transverse slice to the orbit ${\cal O}_\rho$ in the space which is freely generated by adjoint-valued variables. 
   
 Let  the Higgs branch be  $\cal H$ and the Coulomb branch be  $\cal C$, and 
 denote the electric theory by a subscript $e$ and the magnetic theory by a subscript $m$. Then the Higgs branch ${\cal H}_e$ of the electric theory and the Coulomb branch ${\cal C}_m$ of the magnetic theory  are given by the intersection \cite{Bri,Slo}
\beq
{\cal H}_e = {\cal C}_m = {\cal S}_{ \rho} \cap \overline{\cal O}_{\hat\rho^T}\ ,
\eeq
while the Higgs branch ${\cal H}_m$ of the magnetic theory and the Coulomb branch ${\cal C}_e$ of the electric theory are given by
the mirror intersection
\beq
{\cal H}_m = {\cal C}_e = {\cal S}_{\hat\rho} \cap \overline{\cal O}_{\rho^T}.
\eeq
 
 As an example, we can   compute three of the moduli spaces 
for the partitions of  Figure \ref{fig:1} 
by noticing that for the partition $\rho$ in this figure, the Slodowy slice is the whole freely-generated
adjoint-valued  variety, and therefore the Higgs branch of the electric theory becomes the closure of the corresponding nilpotent orbit.
\beq
{\cal H}_e = {\cal C}_m = \overline{\cal O}_{\hat\rho^T}.
\eeq
The latter  have simple descriptions as algebraic varieties with relations given by conditions on matrices.
Explicitly, for models   A, B, C  of Figure \ref{fig:2} 
we have
\begin{subequations}
\bea
{\cal H}_e^A &=& {\cal C}_m^A = \overline{\cal O}_{\hat\rho_A^T} = \left \{M_{8\times 8} : \tr(M) = \tr(M^2) = 0,\, M^3 = 0,\,  rk(M) \le5\right\},\\
{\cal H}_e^B &=& {\cal C}_m^B = \overline{\cal O}_{\hat\rho_B^T} = \left \{M_{8\times 8} : \tr(M) =  \tr(M^2) = 0,\,  M^3 = 0,\,  rk(M) \le4\right\},\\
{\cal H}_e^C &=& {\cal C}_m^C = \overline{\cal O}_{\hat\rho_C^T} = \left \{M_{8\times 8} : \tr(M) =  \tr(M^2) = 0,\,  M^3 = 0,\,  rk(M) \le2\right\}, 
\label{algvar}
\eea
\end{subequations}  
where $rk(M)$ is the rank of the matrix $M$. 

            In the language used in  the previous subsection all  electric quivers in Figure \ref{fig:2} have   two
            hypermultiplets. Call for short  $(q, \tilde q)$ the fields  in the fundamental of $SU(8)$, and $(u, \tilde u)$
            those corresponding to the horizontal link.   The F-flatness conditions are  $\tilde u u = u \tilde u + \tilde q q = 0$.
            It follows that the  meson matrix $M = q \tilde q$  obeys  the conditions $\tr(M) = \tr(M^2) =  M^3 = 0$.
            Furthermore its rank cannot exceed the rank  of the gauge group   under which the quarks $q$ and $\tilde q$
            are charged. This explains the equations (4.7). 
            
  The global symmetry on these branches is $SU(8)$. This is  seen in  the electric quivers as   flavor symmetry, 
and it manifests itself in  the magnetic quivers from the balanced $A_7$ Dynkin diagram\,.\footnote{A
  balanced node is one that has the number of flavors equal to twice the number of colors. 
  For linear or circular quivers this means $2n_a = n_{a{+}1} + n_{a{-}1} + N_a$. {Amusingly this is nothing but the condition for vanishing $\beta$ functions of the  $D=4$ SYM theory with the same quiver.}}  Denoting the
  highest-weight fugacities of $SU(8)$ by $\mu_i$, $i=1,..., 7$ and the highest-weight fugacity of $SU(2)_{H(C)}$ by $t_{1(2)}$, we find
  for  all three  moduli spaces 
  that the operators up to dimension $\Delta = R = 2$  transform in  the following representations
\beq
\label{reps1}
1+\mu_1 \mu_7 t_1^2 + (\mu_1^2 \mu_7^2 + \mu_2 \mu_6 + \mu_1 \mu_7)t_1^4.
\eeq
  At $\Delta = 1$ we find the adjoint of $SU(8)$,  and at $\Delta = 2$ there are  
  1232 + 720 + 63 distinct operators, the same number as in the symmetric traceless product of two
  adjoints.\,\footnote{We already checked  that the trace of the product representation, $\tr M^2$,   
  is indeed zero. }This agrees with our statement that there is  a basis consisting of only open strings.
  
 \smallskip                     
                                                
 The other branches  have the  global symmetry $\bigl( U(1)\times U(2)\bigr)/U(1) \simeq SU(2)\times U(1)$ 
with,  again,  a  universal sector in the adjoint representation  at dimension 1, but different content at higher
 dimensions.  This is clear  from our earlier    enumeration.
 Letting $\mu$ and $\alpha$ be the two fugacities of $SU(2)\times U(1)$,  
 we find the following  content up to dimension 2
\begin{subequations}
\bea
A & : & 1+(\mu^2+1) t_2^2 + \mu (\alpha +\alpha^{-1}) t_2^3 + (\mu^4 + \mu^2 + 1)t_2^4 \ , \\
B & : & 1+(\mu^2+1) t_2^2 + (\mu^4 + \mu^2 + 1)t_2^4 + \mu (\alpha +\alpha^{-1}) t_2^4 \ , \\
C & : & 1+(\mu^2+1) t_2^2 + (\mu^4 + \mu^2 + 1)t_2^4.
\label{reps2}
\eea
 \end{subequations}
 These calculations can be checked from the Coulomb branch of the  electric quivers. 
   In addition to the universal  adjoint, and (adjoint)$^2$ sectors, we have new operators in the 
    bi-fundamental of  $SU(2)\times U(1)$, with $\Delta  = 3/2, 2$ and  $7/2$ for theories A,B and C, as explained. 
    
 \smallskip
 
       For   the conformal manifold   one  must   consider all $\Delta = 2$ operators, including 
       the product of   D5-brane strings  in the adjoint of the electric flavor group with NS5-brane
     strings in the adjoint of the magnetic flavour group.  Notwithstanding additional mixed-branch operators,
     the number of candidate moduli even in these simple models  is formidable\,! Furthermore,  the majority
     of marginal fields  are 2-string bound states which would go undetected  by a  semiclassical analysis on the gravity side. 
  
     The actual  conformal manifold is  the  quotient \eqref{quotient} -- \eqref{quotient1}
      of the space of  all candidate  moduli by the complexified flavor group.  This removes  63+3+1 + 1 = 68  directions,
      i.e. one direction per each generator of the flavor symmetry and one for the accidental $U(1)_F$
      at the enhanced ${\cal N}=4$ supersymmetry point. Denote the space of all $U(1)$ symmetries which are preserved by the conformal deformation by ${\cal P}$. The $U(1)_R$ in the ${\cal N}=2$ superconformal algebra is a particular linear combination of all $U(1)$s in ${\cal P}$. The remaining $U(1)$ charges in ${\cal P}$ can be arranged in a convenient orthogonal basis ${\cal O}$. The conformal manifold is given by the K\"ahler quotient with respect to $U(1)$ symmetries in ${\cal O}$.  
      Notice that  
      at least two charges of opposite sign are necessary to solve 
    $U(1)$   D-term conditions. 
      The particular choice of $U(1)_R$ as the diagonal subgroup of $U(1)_H \times U(1)_C$ implies that there must exist at least one marginal operator from the Higgs branch and 
     one from the Coulomb branch, or else the conformal manifold is trivial.

        To calculate the  manifold one  proceeds  with the methods 
presented in \cite{Benvenuti:2005wi}, with the help of
plethystic techniques as in \cite{Benvenuti:2006qr}.
  To be more specific, one  computes  the set of holomorphic functions from  the
Hilbert series of this moduli space with a  Molien Weyl
integral over the  group of global symmetries. In  the examples at hand, 
both the dimension of the group and the number of marginal operators is large, 
    and the integrals cannot be done explicitly.  Here we will only  sketch  the
    calculation,  which we    illustrate   in appendix \ref{app2}
     for  a simpler case, $U(1)$ theory with one flavor, 
    for which   explicit formulae are available.

               The formula uses the characters for the  representations of the moduli, which  
   can be found in standard references such as Lie Online.
The representations are rather large, hence we will adopt a short-hand notation to denote the expression for the integral.
The  Haar measure for a group $G$  is denoted  $d\mu(G)$, and
the   character  of a representation of  $SU(8)\times SU(2)$ 
     by the  Dynkin labels $[n_1, \ldots, n_7; n]$. 
        With these notational conventions, the
          integral for (the most interesting) theory B  takes the form
\begin{equation}
 \oint{dz\over z}  \oint{dw\over w}  
 \int d\mu_{SU(8)} d\mu_{SU(2)}  \,   {\cal Z}  \nonumber
\end{equation}  \vskip -4mm  where   
  \bea\nonumber
{\cal Z } && \hskip -8mm
 = {\rm PE}   \Bigl[ ([2,0,0,0,0,0,2;0]+[0,1,0,0,0,1,0;0]+[1,0,0,0,0,0,1;0])z^2  q^2 \\ 
  &+& \left. ([\vec{0};4]+
 [\vec{0};2] +  [\vec{0};1](w+ w^{-1}) + [\vec{0};0])z^{-2} q^2  + [1,0,0,0,0,0,1;2] q^2  \right]\ .  
\eea
Here PE  is the plethistic exponential \cite{Benvenuti:2006qr}, 
$w$ is the fugacity of the $U(1)$ magnetic-flavor group, and $z$ is the fugacity of the accidental $U(1)_F$
symmetry which is always present at the   points of moduli space where supersymmetry is  
enhanced to ${\cal N}=4$. 
We have also introduced a fugacity $q$ to  keep  track of the number of couplings in the problem, which can be identified with the residual R symmetry.

  
  \subsection{Mixed-branch operators}   
        
              So far we have considered chiral operators on the Higgs branch where the moduli space is classical,
              or on the Coulomb branch which is   the Higgs branch of the mirror quiver.  These operators organize themselves
              in   $B_1[0]_{2}^{(2;0)}$ and $B_1[0]_{2}^{(0;2)}$  multiplets of $\mathfrak{osp}(4\vert 4)$. 
                   There exist, however,  also mixed-branch operators. The simplest are 
                   products  of a current multiplet of  the electric flavor group with a current multiplet of  the magnetic  flavor group.
                   These  marginal $B_1[0]_{2}^{(1;1)}$ operators are  bound states of a string on the D5-branes with a
                   string on the NS5-branes. They are part of   the  universal spectrum of the CFT.

                     A natural question to ask is whether there exist  
                     single-string   (non-factorizable)  mixed-branch operators.   
                Let us   consider gauge-invariant products of 
                chiral fields which, in addition to the  $q, \tilde q$  coming from  hypermultiplets, also
                involve the   fields  $\Phi_\ell$  sitting in  the ${\cal N}=4$  vector multiplets. 
                            In the  ultraviolet theory, which is free,  the $\Phi_\ell$  belong to the   
              representation   $B_1[0]_{1}^{(0;1)}$ which contains   a triplet of scalars and
              the conserved topological current  $\epsilon^{\mu\nu\rho} F_{ \nu\rho}$.
              Each   $\Phi_\ell$   adds therefore   
              one unit of $SU(2)_C$ spin,  and raises    the scaling dimension by 1.  
              The counting is   valid in the free theory,  but because the ${\cal N}=4$ 
              multiplet  in which such operators sit, is absolutely protected, 
              we may   conjecture  that
              it continues to hold  in the strongly-coupled low-energy theory  as well.

\begin{figure}[t!]
\centering 
 \vskip -1.6 cm
\includegraphics[width=.80\textwidth,scale=0.70,clip=true]{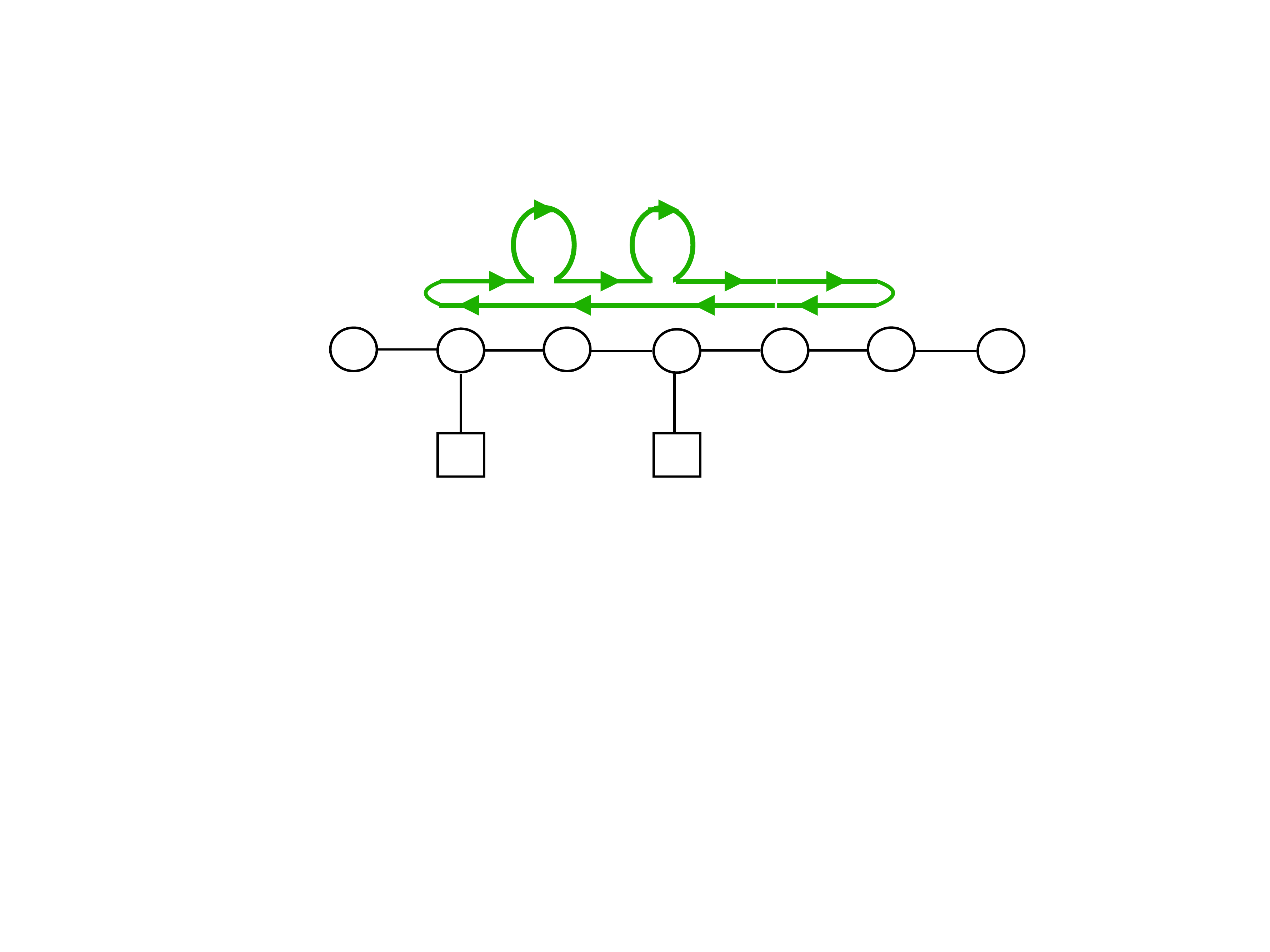}
  \vskip -45mm
   \caption{\small   Mixed-branch operators are gauge-invariant products of    chiral fields from
    both  hypermultiplets
   and   vector multiplets.  The former are   linear string bits 
   while the latter,  in the adjoint representation of the gauge group,  are denoted by
   bubbles. The closed string in the figure has  8 linear segments and 2 bubbles, and transforms
   in the representation $B_1[0]_{6}^{(4;2)}$. }
\label{fig:5} 
 \end{figure}

      Such    operators  can be again  depicted as strings
     which, in addition to line  segments,   contain `bubbles'  at  circular nodes, 
      see  figure \ref{fig:5}\,.  
      Each line  segment stands for a hypermultiplet and   contributes     $\Delta=R = 1/2$ to the operator dimension
      and   the $SU(2)_H$ spin, 
      while  each bubble  stands for   a $\Phi_\ell$ insertion and contributes
      $\Delta=R^\prime  = 1 $ to the operator dimension and the $SU(2)_C$ spin. 
   The resulting   operator  is in  the 
       $B_1[0]_{R+R^\prime}^{(R;R^\prime)}$  multiplet,  with
       $R^\prime$  the
       number of bubbles and 
             $2R$  the number of line segments of the string.

        How many of these operators are independent?  In addition to the conditions  \eqref{Fold} we must now
        also impose  the F-term conditions of $q, \tilde q$  which read 
        \bea
        q_{\square, \ell}\,  \Phi_\ell  = \Phi_\ell\, \tilde q_{\ell, \square}= 0 \ , \qquad 
        q_{\ell+1, \ell}\,  \Phi_\ell \sim   \Phi_{\ell+1}\, q_{\ell+1, \ell}\ ,  \qquad
        \tilde q_{\ell , \ell+1}\,  \Phi_{\ell+1}  \sim   \Phi_{\ell }\, \tilde q_{\ell , \ell+1}\ . 
        \eea
       The `$\sim$' sign in the above relations allows for different gauge couplings of the factor groups. 
       The second and third relations show that bubbles can  move  freely around the  string, 
       while the first equation shows  that
       if they encounter  a vertical segment they vanish.  To avoid this from happening,
        such mixed-branch single-string   operators  must thus be closed. 
       This   fits nicely  with the   analysis in   section  \ref{3.1} 
      where we argued  that   $B_1[0]_{R+R^\prime}^{(R;R^\prime)}$ multiplets  with $RR^\prime \not= 0$ 
       are either multiparticle states or 
         Kaluza-Klein descendants of the  gravitini. 
   
  \smallskip 
      Such single-closed-string mixed-branch  operators exist in theories based on circular quivers  (whose
      dual supergravity backgrounds were found  in \cite{Assel:2012cj}). 
     For  linear quivers  they  vanish by a similar    argument as the one used in section \ref{4.3}.  As explained there, 
        one can  `fold and slide'  the closed string until it  reaches
       the boundary of the linear quiver,  where it is annihilated.  It can be seen that the bubble  insertions go along for the ride,
       their role being only to annihilate any open strings emitted in the process.

   \vskip 6mm
   
  {\bf Aknowledgements}:   We have benefited from discussions with
    Antonio Amariti, Santiago Cabrera, Francesco Fucito, Costas Kounnas, Ioannis Lavdas, 
    Bruno Le Floch,  Jan Louis, 
    Severin L\"ust, Noppadol Mekareeya, Francisco Morales,
      Marios Petropoulos, Raffaele Savelli, Nathan Seiberg, Tin Sulejmanpasic and  Hagen Triendl. 
 {Part of the work was carried on during the INdAM - INFN program {\it ``Mathematics and Physics at the Crossroads}. We thank the organisers and the participants for creating a stimulating environment.}
   M.~B. would like to thank the MIUR-PRIN contract 2015MP2CX4002 {\it ``Non-perturbative aspects of gauge theories and strings''} for partial support.


      \vskip 1cm  

   \appendix
\section{The $\boldsymbol{{\cal N}=4}$ supergravity
solutions}\label{app1}
        
    The local form of the most general  Type IIB  
     solutions  with  $\mathfrak{osp}(4\vert 4)$ symmetry
      was derived in \cite{D'Hoker:2007xy, D'Hoker:2007xz}. 
It depends on two real non-negative functions $h_1$ and $h_2$
that are
harmonic on the surface $\Sigma$. The coordinates on this latter
are $(z, \bar z)$.
Changing  slightly the  notation of  the above references 
 we  define  the  auxiliary functions 
  \bea\label{a1}
W = \partial_z \partial_{\bar{z}} (h_1h_2)\qquad {\rm and} \ \ \
\ { \cal U}_i = 2 h_1h_2|\partial_z h_i|^2 - h_i^2 W\ .
   \eea 
The 10-dimensional manifold is a fibration of
AdS$_4\times$S$^{2}_{(1)}\times$S$^{2}_{(2)}$
 over  $\Sigma$ with metric
   \bea\label{first}
{4\over \alpha^\prime}\, 
ds^2 = \rho_4^2\, ds^2_{\rm AdS} + \rho_1^2\, ds^2_{{(1)}} +
\rho_2^2\, ds^2_{ {(2)}} +
4 \rho^2\, dzd\bar{z} \ , 
\eea
\vskip 1mm  
\noindent where  $\alpha^\prime$ is the Regge slope, 
$ds^2_{ {(i)}} = d\vartheta_i^2 + \sin\vartheta_i^2
d\varphi_i^2\ $ are the metrics of the unit-radius
2-spheres (also called S$^2$ and S$^{2\,\prime}$ in the main
text)
and the  scale  factors  read
\bea
\rho_1^8 = 16 h_1^8 \,{ {\cal U}_2 W^2 \over {\cal U}_1^3} \ ,
 \qquad
\rho_2^8 = 16 h_2^8\, { {\cal U}_1 W^2 \over {\cal U}_2^3} \ ,\eea
 \bea
\rho^8 = {{ \cal U}_1  {\cal U}_2 W^2 \over h_1^4 h_2^4} \ , 
 \qquad
\rho_4^8 = 16\, {{ \cal U}_1  {\cal U}_2\over  W^2} \ .   
\eea
   The solutions  have  a non-trivial dilaton field
\bea
e^{2\phi} = \sqrt{{\cal U}_2 \over {\cal U}_1}\ , 
\eea
as well as antisymmetric Neveu-Schwarz and Ramond-Ramond 3-form
fields
 \bea
{4\over \alpha^\prime}\, {\rm Re}(f_3) = \omega^{(1)} \wedge
d\beta_1\qquad {\rm and}\ \ \ \ \
{4\over \alpha^\prime}\, {\rm Im}(f_3) = \omega^{(2)} \wedge
d\beta_2
 \eea
with
$\omega^{(i)} = d\cos\vartheta_i\wedge d\varphi_i
$
  the  2-sphere volume forms and  
 \bea
\beta_1 = 2i \, {h_1\over {\cal U}_1} h_1h_2 (\partial_z h_1
\partial_{\bar{z}} h_2 - \partial_z h_2 \partial_{\bar{z}} h_1)
+ 2 \widetilde{h}_2\ ,
 \eea
  \bea
\beta_2 = 2i \, {h_2\over {\cal U}_2} h_1h_2 (\partial_z h_1
\partial_{\bar{z}} h_2 - \partial_z h_2 \partial_{\bar{z}} h_1)
- 2 \widetilde{h}_1\ .
 \eea
Here $\tilde h_i$ are the functions dual to $h_i$, i.e. in terms
of two meromorphic functions ${\cal A}_i$ :
\bea
h_1 = -i(A_1 - \bar{A}_1)\ , \quad h_2 = A_2 + \bar{A}_2 \ ,
\nonumber
\eea
\bea
\widetilde{h}_1 = A_1 + \bar{A}_1\ , \quad \widetilde{h}_2 =
i(A_2 - \bar{A}_2)\ .
\eea
Finally, there is a non-trivial self-dual Ramond-Ramond  5-form  \bea\label{f5}
 ({4\over \alpha^\prime})^2\, 
 f_5  =   - 4 \rho_4^4 \, \omega^{\rm AdS} \wedge {\cal F} 
+ 4 \rho_1^2 \rho_2^2\, \omega^{(1)} \wedge \omega^{(2) } \wedge
(*_{2} {\cal F}) \ ,
\eea
where $\omega^{\rm AdS}$ is the volume form of the unit-radius
AdS$_4$, and $\rho_4^4 \, {\cal F} $
is a closed one-form given by the following formidable-looking
expressions:
 \bea
{\cal F}  =   \rho_4^{-4}\, d{\cal J} \ , \quad {\rm with}\quad\quad {\cal J} = 3({\cal C} + \bar {\cal C} ) - 3 {\cal D} + i
{h_1h_2 \over W} (\partial_z h_1 \partial_{\bar{z}} h_2 -
\partial_z h_2 \partial_{\bar{z}} h_1) \ , \nonumber
\eea
 \bea\label{last}
\partial{\cal C} = A_1 \partial_z A_2 - A_2 \partial_z A_1\ ,
\quad {\rm and}\quad {\cal D} = A_1 \bar{A}_2 + A_2 \bar{A}_1\ .
\eea
The  1-form \ $*_{2} {\cal F} $ 
  in \eqref{f5}
is  the  dual  of  ${\cal F} $ on the Riemann surface $\Sigma$.  
 
     
   The explicit solutions found in \cite{Assel:2011xz} 
 correspond to the following choices  for the
meromorphic functions $A_i$ (here rescaled by a factor
$4/\alpha^\prime$)
\bea \label{strip}
 i A_1 = \sum_{a=1}^p N_a \log\left(   
 {1 +  i e^{z-\delta_a}\over 1 -i  e^{ z - \delta_a } }
  \right)   \ ,   
\qquad
 A_2 =  \sum_{{\hat a}=1}^{\hat p} \hat N_{\hat a} \log \left(   
{1 +  e^{-z+ \hat \delta_{\hat a}}\over 1 - e^{ -z + \hat \delta_{\hat a} } }  \right)  \ , 
\eea
where $z$ parametrizes the infinite strip $0\leq {\rm Im} z \leq
\pi/2$.
These solutions are holographically dual to linear-quiver gauge
theories at their infrared fixed points.
To find the supergravity backgrounds for
circular quivers one simply sums over an infinite array, $A_i
\to \sum_n A_i(z+nt)$,
before identifying $z$ periodically ($z \equiv z + t$)
\cite{Assel:2012cj}.

The function $A_1$ in \eqref{strip} has logarithmic
singularities at $ z= \delta_a + {i\pi/2}$ on the upper strip
boundary, while $A_2$
has logarithmic singularities  at
$ z= \hat\delta_{\hat a}$ on the lower boundary. These are the
locations, respectively, of (stacks of $ N_a$) D5-branes
and (stacks of $\hat N_{\hat a}$) NS5-branes. Both  the coefficients $\{ N_a, \hat N_{\hat a}\}$ and the 
 positions  $\{\delta_a, \hat\delta_{\hat a}\}$
 of these
singularities  on the boundary  are continuous parameters of the supergravity solution,
 but in string theory they are quantized. 
 This is because the former are five-brane charges and the latter 
are related to D3-brane charges of the five-brane stacks by the following equations \cite{Assel:2011xz}
\bea
\ell_a = -   \sum_{{\hat a}=1}^{\hat p}\hat N_{\hat a}\, {2\over \pi} \arctan
(e^{- \delta_a+ \hat\delta_{\hat a} })
  = - {i\over \pi}
A_2(z=\delta_a + {i\pi\over 2})\ ,
 \eea
\bea
\hat \ell_{\hat a} =   \sum_{a=1}^{ p} N_a \, {2\over \pi} \arctan
(e^{- \delta_a+ \hat\delta_{\hat a} })
  = {1
\over \pi} A_1(z=\hat \delta_{\hat a})\ ,
\eea
where $\ell_a$ is the D3-brane  charge  (or linking number) of a D5-brane in the 
$a$th  stack and
$\hat \ell_{\hat a}$  is the D3-brane  charge of a  NS5-brane in the $\hat a$th stack.
Note that  one equation is trivial, because the total D3-brane charge is zero, 
reflecting the arbitrariness in  choosing   the origin of the Re$z$ axis. 
 The remaining  $(p+\hat p -1)$ equations  
determine the $(p+\hat p-1)$   5-brane positions.  
 So  the solutions \eqref{strip} have no continuous  moduli.
\smallskip

     Near a NS5-brane singularity,  $z- \hat\delta_{\hat a} = r e^{i\theta}$
     ($r\ll 1$),    the  meromorphic functions read
     \bea\label{sing}
    A_1 = \pi {\hat \ell_{\hat a} } +  c_1\,   r e^{i\theta}  + O(r^2) \ , \qquad 
    A_2 =  - \hat N_{\hat a} \log (r e^{i\theta})  + c_2  + O(r)\ , 
     \eea
where  the expansion coefficients $c_1, c_2$ are both real,
\bea
c_1 = \sum_{a=1}^{ p} \, {N_a\over \cosh (\delta_a - \hat\delta_{\hat a})
}\ , \qquad
c_2 = \hat N_{\hat a} \log 2 + \sum_{\hat b \not= {\hat a}} \hat N_{\hat b }
\log \left(
{1 + e^{\hat\delta_{\hat b} - \hat \delta_{\hat a}}\over 1 -
e^{\hat\delta_{\hat b} - \hat \delta_{\hat a}} }
  \right) \ . 
\eea
Inserting  \eqref{sing} in  \eqref{a1}-\eqref{last}  gives the  following singular metric and dilaton:  \vskip -2mm
 \bea
 ds^2 \,  \simeq \,   \alpha^\prime c_1^2 \hat N_{\hat a}^2\,   (- {\log r\over r^2})^{-{1/ 4}}\,\Bigl( {dr^2\over r^2}  +   \, 
  ds^2_{{\rm S}^3}+    (-\log r)  [ ds^2_{\rm AdS} + ds^2_{(2)}]   \Bigr) 
\eea
\bea
e^{2\phi} \,  \simeq \,  {\hat N_{\hat a}^2\over  c_1^2}\,  (- {\log r\over r^2})^{1/2}   
\eea
Note in particular that the singularity is at finite distance since the integral $\int {dr \over r} (- {\log r\over r^2})^{-1/8}$ converges at $r\simeq 0$.   Similar expressions hold for the D5-brane singularities.



\section{Conformal manifold of $U(1)$ with 1 flavor}
\label{app2}

The  simplest  set of examples are those conformal
manifolds which have points on their moduli spaces with free
theories, or more generally, with BPS spectra that are generated
by free operators.  Although these theories correspond to singular supergravity backgrounds,
the study of their moduli demonstrates the method outlined in  section \ref{4.3}.

Consider the case   of SQED which is well known to have
a one (quaternionic) dimensional Coulomb branch and no Higgs
branch.\,\footnote{This means that there can be  no conformal manifold corresponding to the
maximal embedding \ref{pick}, but there is one if we  pick  the ${\cal N}=2$  R-symmetry to lie
entirely inside $SU(2)_C$.
}
Details of this theory are conveniently summarized in
\cite{Cremonesi:2013lqa} and are presented below. The theory
starts as an interacting theory in the UV and flows to a free
theory in the IR. Even though the theory is free in the IR, the
free operators are in fact non perturbative objects which carry
a magnetic charge. The Coulomb branch is a copy of $\IH=\IC^2$,
and the ring of holomorphic BPS operators is the set
\beq
O_{m,k} = V_m \phi^k, \qquad m\in\IZ, \quad k=0, 1, 2, \ldots
\qquad \Delta(m,k) = \frac{1}{2}|m|+k
\eeq
with $V_m$ a bare monopole operator of magnetic charge $m$ and
conformal dimension $\Delta(m) = \frac{1}{2} |m|$ and $\phi$ is
a complex scalar in the vector multiplet with conformal
dimension 1. $\Delta(m,k)$ is the conformal dimension of
$O_{m,k} $. The ring is freely generated, as the moduli space is
$\IH$ and this implies non trivial relations that these
operators satisfy. They can be written as
\beq
O_{m,k} \,O_{m',k'} =
O_{m+m',k+k'}\,\phi^{\Delta(m)+\Delta(m')-\Delta(m+m')}
\eeq
or perhaps a slightly simpler expression for bare monopole
operators,
\beq
V_{m} \,V_{m'} =
V_{m+m'}\,\phi^{\Delta(m)+\Delta(m')-\Delta(m+m')}
\eeq
from which one can derive relations like
\beq
V_m = (V_1)^m, \quad V_{-m}=(V_{-1})^m, \qquad m>0
\eeq
and perhaps the most crucial relation
\beq
V_1 V_{-1} = \phi
\eeq
where $V_1$ and $V_{-1}$ are the generators of the ring.
The global symmetry of this theory is $Sp(1)=SU(2)$ under which the
monopole operators $V_1$ and $V_{-1}$ transform in the
fundamental (doublet) representation.
The set of holomorphic BPS operators can be recast into a
collection of representations of $Sp(1)$ with fixed conformal
dimension. We have a spin $j$ representation of $Sp(1)$ for every
operator of conformal dimension $j$.
This can be summarized by the highest weight generating function
(HWG) \cite{Hanany:2014dia}
\beq
\label{HWG1}
\frac{1}{1-\mu t}
\eeq
where $t$ is the fugacity for the conformal dimension (or
alternatively the fugacity for $SU(2)_C$)
and $\mu$ is the fugacity for the highest weight $n=2j$ of the
representation with spin $j$.
 For example we list the operators up to spin 2 of $Sp(1)$.
  \bea  \nonumber
   V_0 &0& \\  \nonumber
 V_1, V_{-1} &\frac{1}{2}& \\ \nonumber
 V_2, \phi, V_{-2} &1& \\ \nonumber
 V_3, V_1 \phi, V_{-1} \phi, V_{-3} &\frac{3}{2}& \\ \nonumber
 V_4, V_2 \phi, \phi^2, V_{-2} \phi, V_{-4} &2&
    \eea
At conformal dimension 2 we find 5 different operators which can
be added with arbitrary coefficients $\lambda_{ij}, i,j=1,2,3$ symmetric such that $\lambda_{ii}=0$ as marginal
deformations, breaking supersymmetry from ${\cal N}=4$ to ${\cal
N}=2$. Recalling that the moduli space is $\IC^2$ with a $U(2) =
U(1)_r \times SU(2)$ isometry that preserves ${\cal N}=2$
supersymmetry, where $U(1)_r$ is the ${\cal N}=2$ R symmetry, we
find that the $SU(2)$ symmetry rotates the 5 parameters and
leaves a 2 complex dimensional conformal manifold.
To compute this manifold we proceed with the methods that are
presented in \cite{Benvenuti:2005wi}, with the help of
plethystic techniques as in \cite{Benvenuti:2006qr}. The moduli
space of conformal deformation is given by the K\"ahler quotient
of the gauge theory of $SU(2)$ coupled to 1 chiral multiplet in
the spin 2 representation.
To compute the set of holomorphic functions we compute the
Hilbert series of this moduli space with the Molien Weyl
integral
\beq
\oint_{|z|=1} dz \frac{1-z^2}{2\pi i z} {\rm PE}\left [ (z^4 + z^2 + 1 +
z^{-2} + z^{-4}) q \right ] = \frac{1}{(1-q^2)(1-q^{3})}
\eeq
where the fugacity $q$ measures the number of $\lambda$'s, to be
consistent with equation (\ref{HWG1}).
This computation has the interpretation that the conformal
manifold is a copy of $\IC^2$ and is freely generated by two
invariants which are quadratic and cubic in $\lambda$, say $c_2 = \lambda_{ij}\lambda_{ij}$
and $c_3 = \lambda_{ij} \lambda_{jk} \lambda_{ki}$. The functions $c$ are globally defined on the
conformal manifold as they are invariant under the $SU(2)$
rotations. These invariants can be used to construct any
globally defined function on the conformal manifold, and in
particular any physical quantity on the conformal manifold must
be a function of these two invariants.



 \end{document}